# LS-DYNA Machine Learning-based Multiscale Method for Nonlinear Modeling of Short Fiber-Reinforced Composites

Haoyan Wei[1*✉], C. T. Wu[1], Wei Hu[1], Tung-Huan Su[1],

Hitoshi Oura[2], Masato Nishi[2],

Tadashi Naito[3],

Stan Chung[4], Leo Shen[4]

[1]ANSYS Inc.,
*7374 Las Positas Rd, Livermore, California, 94551, USA*

[2]JSOL Corporation,
*Kudan-Kaikan Terrace 1-6-5, Kudanminami, Chiyoda-ku, Tokyo, 102-0074, Japan*

[3]Honda Motor Co., Ltd.,
*4630 Shimotakanezawa, Haga-machi, Haga-gun, Tochigi, 321-3393, Japan*

[4]CoreTech System Co., Ltd.,
*8F-2, No.32, Taiyuan St., Zhubei City, Hsinchu County 302, Taiwan*

*✉  Corresponding Author's Email: haoyan.wei@ansys.com*

## Abstract

*Short-fiber-reinforced composites (SFRC) are high-performance engineering materials for lightweight structural applications in the automotive and electronics industries. Typically, SFRC structures are manufactured by injection molding, which induces heterogeneous microstructures, and the resulting nonlinear anisotropic behaviors are challenging to predict by conventional micromechanical analyses. In this work, we present a machine learning-based multiscale method by integrating injection molding-induced microstructures, material homogenization, and Deep Material Network (DMN) in the finite element simulation software LS-DYNA for structural analysis of SFRC. DMN is a physics-embedded machine learning model that learns the microscale material morphologies hidden in representative volume elements of composites through offline training. By coupling DMN with finite elements, we have developed a highly accurate and efficient data-driven approach, which predicts nonlinear behaviors of composite materials and structures at a computational speed orders-of-magnitude faster than the high-fidelity direct numerical simulation. To model industrial-scale SFRC products, transfer learning is utilized to generate a unified DMN database, which effectively captures the effects of injection molding-induced fiber orientations and volume fractions on the overall composite properties. Numerical examples are presented to demonstrate the promising performance of this LS-DYNA machine learning-based multiscale method for SFRC modeling.*

## Introduction

Short-fiber-reinforced composites (SFRC), such as glass-fiber-reinforced thermoplastics, become increasingly attractive for lightweight structural applications in automotive and electronics industries. Nowadays, mass-production of SFRC parts is achieved by the injection molding process, which inevitably causes inhomogeneous fiber dispersion in the polymer melt, and consequently, location-dependent fiber orientations and volume fractions in the finished products. The heterogeneous microstructural distributions and the distinct properties of each constituent material lead to highly complicated, anisotropic, and nonlinear responses of SFRC (*Mortazavian and Fatemi 2015; Hessman et al. 2019*), and reliable prediction of the mechanical behaviors of composite products remains challenging. Phenomenological anisotropic constitutive models often require a tedious parameter fitting process to calibrate a large number of model parameters against material data. Measuring these material data from a series of physical experiments is quite time-consuming, and often the fitted model parameters become ineffective for a different microstructure in materials. For instance, conventional constitutive models calibrated for SFRC with low fiber volume fraction is not able to predict SFRC with high fiber volume fraction, even if the fiber orientations remain unchanged. To circumvent such difficulties, multiscale composite material modeling methods have emerged as an effective means to predict macroscopic material properties at the upper length scale from the geometries and properties of the materials at a lower length scale.

Various multiscale methods have been developed for modeling composites. Among these methods, analytical homogenization methods (*Eshelby and Peierls 1957; Mori and Tanaka 1973; Nemat-Nasser and Hori 2013; Li and Wang 2008; Huang 2021*) are quite popular due to their simplicity and efficiency. For instance, upscaling microscopic material information to obtain homogenized SFRC material properties via analytical micromechanics methods has been studied in *(Müller and Böhlke 2016; Tucker III and Liang 1999*). Nevertheless, analytical homogenization methods are based on assumptions involving simplified microstructural morphologies and material behaviors, which limits their applicability and accuracy for nonlinear analysis of composites with complicated microstructures. On the other hand, computational homogenization methods become increasing attractive in multiscale design and modeling of composite materials *(Fish et al. 2021; Terada et al. 2013*). For SFRC, Representative Volume Element (RVE) with realistic fiber distributions can be numerically reconstructed, and multiscale mechanical responses can be predicted through Direct Numerical Simulations (DNS) methods, such as the Finite Element Method (FEM) and Fast Fourier Transformation (FFT) *(Naili et al. 2020; Müller et al. 2015*). The high-fidelity DNS is especially useful for the design and analysis of composite materials at the RVE level. To model large-scale composite structures, multiscale methods coupling numerical RVE models with structure-level FE models have been developed. While FEM is typically adopted to discretize the macroscale composite structure, different numerical approximation methods have been utilized for the microscale RVE models, and the resulting multiscale methods are referred to as FE[2] *(Feyel 2003; Feyel and Chaboche 2000; Kouznetsova et al. 2004; Tan et al. 2020)* or FE-FFT *(Kochmann et al. 2018; Spahn et al. 2014)*. Although these multiscale methods are particularly advantageous to high-fidelity structural analysis, their applications to industrial scale modeling are limited by the high computational costs *(Liu et al., 2020; Xu et al., 2020)*. For large-scale injection-molded SFRC structures with heterogeneous fiber distributions, high-fidelity FE[2] or FE-FFT models will consume extremely high CPU time and memory that are unaffordable especially when nonlinear analyses are desired.

Recent progress in machine learning and data science have brought great opportunities to develop advanced data-driven material modeling and multiscale simulation methods (*LeCun et al. 2015; Goodfellow et al. 2016; Liu et al. 2021; Bishara et al. 2022; Vu-Quoc and Humer 2022*). To circumvent the limitations of conventional constitutive modeling, the model-free data-driven approach has been developed, which formulates an optimization problem to search for a stress solution directly from the material database characterizing constitutive behaviors subjected to essential physical constraints, such as equilibrium and compatibility conditions (*Kirchdoerfer and Ortiz 2016; Ibanez et al. 2018; Eggersmann et al. 2019; He and Chen 2020; He et al. 2020; He et al. 2021a; He et al. 2021b; Xu et al. 2020*). This data-driven computing paradigm has been applied for multiscale modeling of fiber-reinforced plastic composites (*Huang et al. 2021*), biological materials (*Mora-Macías et al. 2020; Sanz-Herrera et al. 2021*), and granular materials (*Karapiperis et al. 2020*). However, its application to elastoplastic material modeling remains challenging due to difficulties in defining a material database to characterize path-dependent material behaviors. Meanwhile, machine learning techniques have been applied to construct surrogate models of constitutive laws, including Gaussian process modeling (*Bostanabad et al. 2018; Chen et al. 2018*) and artificial neural networks, such as feedforward neural networks *(Ghaboussi et al. 1991; Fritzen et al. 2019; Le et al. 2015; Lu et al. 2019)*, recurrent neural networks *(Ghavamian and Simone 2019; Wang and Sun 2018)*, and graph/convolutional neural networks *(Frankel et al. 2019; Vlassis et al. 2020; Rao and Liu 2020)*. In addition, neural network-based constitutive models with embedded material physical constraints including material frame invariance (*Ling and Jones et al. 2016*), symmetric positive definiteness (*Xu and Huang et al. 2021*), self-consistency (*Bonatti and Mohr 2022*), and thermodynamics (*Vlassis and Sun 2021; Masi and Stefanou et al. 2021; He and Chen 2022*) have been developed. Recently, coupling of neural networks with finite element methods and meshfree methods for modeling material damage and strain localization phenomena have also been investigated (*Tao et al. 2022; Baek et al. 2022*). These studies demonstrate the excellent performance of machine learning methods for modeling complex material physics by exploitation of material data.

To accelerate multiscale material modeling, several reduced-order modeling methods have been developed through the construction of surrogate models for high-fidelity numerical simulations. Along this line, model-order reduction based on proper-orthogonal decomposition is developed in (*Kaneko et al. 2021; Rocha et al. 2020; Fritzen and Kunc 2018; Goury et al. 2016; Yvonnet and He 2007*), self-consistent clustering analysis is developed in (*Gao et al. 2020; Liu et al. 2016; Liu et al. 2018; Yu et al. 2019*), and a mechanistic machine learning method named Deep Material Network (DMN) is proposed by (*Liu et al. 2019a; Liu and Wu 2019*). DMN is designed to capture nonlinear microstructural interactions through a binary-tree network structure equipped with physics-based building blocks (*Liu et al. 2019a; Liu and Wu 2019; Gajek et al. 2020*). DMN can be trained in an offline stage to learn the microscopic material morphologies and physics hidden in linear composite material data, and afterwards the trained network is able to perform multiscale online prediction of nonlinear constitutive behaviors. DMN has been extended to model woven composites (*Wu et al. 2021*), porous materials (*Nguyen and Noels 2022a; Nguyen and Noels 2022b*), cohesive interfacial failure *(Liu 2020*), and strain localization analysis (*Liu 2021*). In addition, transfer learning strategies are developed in (*Liu et al. 2019b; Liu et al. 2020; Huang et al. 2022*) for fast creation of DMN models with minimized training efforts for materials that share similar microstructural morphologies but own different characteristic geometries, such as particle-reinforced composites with different volume fractions. Coupling of DMN to finite elements has also been investigated for multiscale structural simulation *(Gajek et al. 2022; Gajek et al. 2021; Liu et al. 2020*). Despite the great progress, the usage of mechanistic machine learning techniques for Computer-Aided Engineering (CAE) is still limited within academic research community due to the lack of a general and robust software platform. To bridge this gap, we have developed a unified DMN database for SFRC to cover a full range of injection-molded microstructures, and the trained DMN model is seamlessly integrated in the multiphysics simulation software LS-DYNA for nonlinear multiscale modeling.

The main goal of this paper is to present the LS-DYNA machine learning-based multiscale method for nonlinear modeling of SFRC, and the remainder of this paper is organized as follows. Firstly, an overview of network architecture of DMN is given. Next, we present the details on the integration of DMN into LS-DNA for SFRC modeling, including the offline training of DMN based on an efficient transfer learning scheme for SFRC, the fast online generation of new network parameters based on injection-molding-induced microstructures, and the nonlinear multiscale prediction of SFRC parts by coupling trained DMN models with finite elements. Lastly, numerical examples are presented to demonstrate the capability of the proposed method, followed by the conclusions.

**Overview of Deep Material Network**

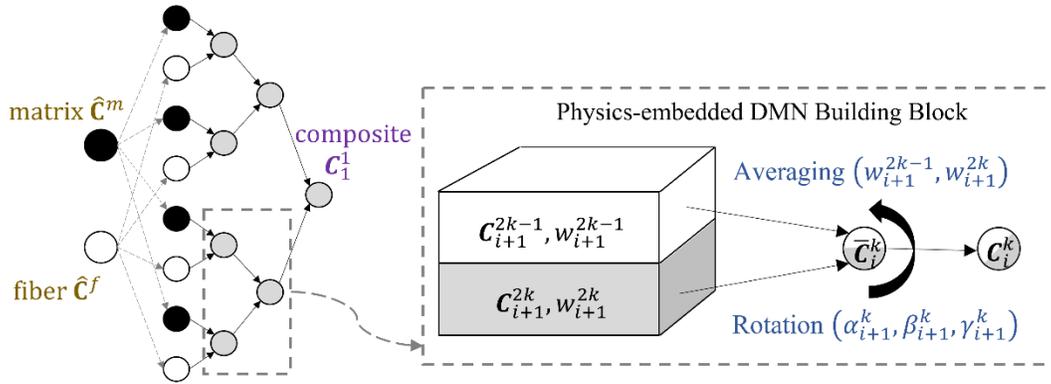

**Fig. 1.** Architecture of a 4-layer Deep Material Network (DMN).

Deep Material Network (DMN) proposed in (*Liu et al. 2019a; Liu and Wu 2019*) is a mechanistic machine learning method for data-driven multiscale material modeling. As illustrated in Fig. 1, a binary-tree network structure is adopted for DMN. Therefore, for a network with $N$ layers, there are $(2^N - 1)$ nodes, where $2^{N-1}$ nodes are located at the bottom layer $N$. For the $k$-th node at layer $i$, four network parameters are defined, including one nodal weight $w_i^k$ and three Euler angles $\alpha_i^k$, $\beta_i^k$, and $\gamma_i^k$, where $1 \leq i \leq N$ denotes the layer index, and $1 \leq k \leq 2^{i-1}$ denotes the node index within each layer. Different from conventional artificial neural networks, all the network parameters of DMN have clear physical meanings, and thus it is straightforward to apply the rule of mixture to calculate an averaged stress $\overline{\sigma}_i^k$ using nodal weights:

$$\overline{\sigma}_i^k = \frac{w_{i+1}^{2k-1}}{w_{i+1}^{2k-1} + w_{i+1}^{2k}} \sigma_{i+1}^{2k-1} + \frac{w_{i+1}^{2k}}{w_{i+1}^{2k-1} + w_{i+1}^{2k}} \sigma_{i+1}^{2k} \tag{1}$$

in which $\sigma_{i+1}^{2k-1}$ and $\sigma_{i+1}^{2k}$ are the stresses associated with the two child nodes of the $k$-th node at layer $i$, $w_{i+1}^{2k-1}$ and $w_{i+1}^{2k}$ are the nodal weights of these two child nodes, respectively. Accordingly, an averaged material stiffness $\overline{C}_i^k$ can be expressed as a function of the nodal weights and material stiffness of its two child nodes:

$$\overline{C}_i^k = C_{i+1}^{2k} - \frac{w_{i+1}^{2k-1}}{w_{i+1}^{2k-1} + w_{i+1}^{2k}} \left( C_{i+1}^{2k} - C_{i+1}^{2k-1} \right) \mathbf{A} \tag{2}$$

where $\mathbf{A}$ denotes a strain concentration matrix within each DMN building block, which is obtained by enforcing the interfacial equilibrium and kinematic conditions of a two-phase composite material (Liu and Wu, 2019). Derivation of an analytical form for the strain concentration matrix is given in ***Appendix I*** of this paper.

In addition to nodal weights, Euler angles $\alpha_i^k$, $\beta_i^k$, and $\gamma_i^k$ are defined at each node of the network to capture directional material behaviors due to complicated microstructures.

By applying three dimensional rotations to the averaged stiffness and stress, one obtains the following rotated stiffness matrix and stress vector:

$$\boldsymbol{C}_i^k = \boldsymbol{R}^T\big(\alpha_i^k, \beta_i^k, \gamma_i^k\big)\overline{\boldsymbol{C}}_i^k\ \boldsymbol{R}\big(\alpha_i^k, \beta_i^k, \gamma_i^k\big) \tag{3}$$

$$\boldsymbol{\sigma}_i^k = \boldsymbol{R}\big(\alpha_i^k, \beta_i^k, \gamma_i^k\big)\ \overline{\boldsymbol{\sigma}}_i^k \tag{4}$$

where $\boldsymbol{R}$ denotes a rotation matrix based on the Euler angles $\big(\alpha_i^k, \beta_i^k, \gamma_i^k\big)$. By applying the averaging and rotation operations to the physical quantities (stress, strain, stiffness, etc.) in different building blocks, DMN can upscale the microscopic base material behaviors at the bottom layer to predict macroscopic composite behaviors at the top layer.

The nodal weight $w_i^k$ of a parent node is computed by adding up the weights of its two child nodes

$$w_i^k = w_{i+1}^{2k-1} + w_{i+1}^{2k} \tag{5}$$

except for the bottom layer nodes whose weights are activated through the rectified linear unit (ReLU):

$$w_N^k = Re(z^k) = max(z^k, 0) \tag{6}$$

Therefore, all the nodal weights in DMN can be calculated from the activations $z^k$. As a result, for a network with $N$ layers, its independent network trainable parameters are $2^{N-1}$ activations $z^k$ and $3(2^N - 1)$ rotation angles $\alpha_i^k$, $\beta_i^k$, and $\gamma_i^k$, where $1 \leq i \leq N$, $\ 1 \leq k \leq 2^{N-1}$. Once DMN trainable parameters are determined through offline training, the essential microstructural interactions can be learned by the trained network, which can be applied for online prediction of nonlinear composite behaviors under general loading conditions.

## LS-DYNA Machine Learning-based Multiscale Method

### Offline Training of DMN for SFRC

Rewriting the independent network trainable parameters in a vector form as $\boldsymbol{z} \in \mathbb{R}^{2^{N-1}}$, $\boldsymbol{\alpha} \in \mathbb{R}^{2^{N-1}}$ , $\boldsymbol{\beta} \in \mathbb{R}^{2^{N-1}}$ , $\boldsymbol{\gamma} \in \mathbb{R}^{2^{N-1}}$ , we can express the overall material stiffness $\boldsymbol{C}_1^1$ of a two-phase composite at the top node of a $N$-layer DMN as:

$$\boldsymbol{C}_1^1 = \boldsymbol{f}\big(\widehat{\boldsymbol{C}}^f, \widehat{\boldsymbol{C}}^m, \boldsymbol{z}, \boldsymbol{\alpha}, \boldsymbol{\beta}, \boldsymbol{\gamma}\big) \tag{7}$$

in which $\widehat{\boldsymbol{C}}^f$ and $\widehat{\boldsymbol{C}}^m$ represent stiffness matrices of the fiber phase and the matrix phase of short-fiber-reinforced composites. During the offline training process, they are assigned to DMN's bottom layer nodes as follows:

$$\boldsymbol{C}_N^k = \begin{cases} \widehat{\boldsymbol{C}}^f, & \text{if } k \text{ is even} \\ \widehat{\boldsymbol{C}}^m, & \text{if } k \text{ is odd} \end{cases} \tag{8}$$

To determine the network trainable parameters, an optimization problem is formulated based on the mean square error (MSE), and the cost function is given by (Liu and Wu 2019):

$$J(\boldsymbol{z}, \boldsymbol{\alpha}, \boldsymbol{\beta}, \boldsymbol{\gamma}) = \frac{1}{2N_s} \sum_{j=1}^{N_s} \frac{\| f(\widehat{\boldsymbol{C}}_j^f, \widehat{\boldsymbol{C}}_j^m, \boldsymbol{z}, \boldsymbol{\alpha}, \boldsymbol{\beta}, \boldsymbol{\gamma}) - \widehat{\boldsymbol{C}}_j^c \|^2}{\| \widehat{\boldsymbol{C}}_j^c \|^2} + \lambda \left( \sum_{k=1}^{2^{N-1}} Re(z^k) - 2^{N-2} \right)^2 \tag{9}$$

where $\| \cdots \|$ denotes the Frobenius norm, $\lambda$ is a positive hyper-parameter associated with the regularization term, which is set to be 0.001 in the present study to ensure the well-posedness of the optimization problem, $j$ denotes the index of the material sample in the training dataset, and $N_s$ is the total number of material samples. Hence, $\widehat{\boldsymbol{C}}_j^f$, $\widehat{\boldsymbol{C}}_j^m$, $\widehat{\boldsymbol{C}}_j^c$ represent the fiber stiffness, matrix stiffness, and composite stiffness of the $j^{\text{th}}$ material sample, all of which are considered as linear elastic material properties. To minimize the cost function, the mini-batch gradient descent algorithm is employed, where gradients of the cost function with respect to the trainable parameters $\nabla J$ are derived by the backpropagation algorithm, as analytical functions are available in DMN building blocks.

Offline training data for DMN, i.e., linear elastic macroscopic stiffness tensors of the composites and microscopic stiffness tensors of the material constituents, can be gathered from both experimental measurements and numerical predictions. In the present study, high-fidelity computational homogenization of SFRC is employed to generate training data due to the lack of experimental data. To this end, RVE models are reconstructed based on SFRC microstructures. In practice, SFRC products may contain heterogeneous microstructures due to numerous combinations of fiber orientations and fiber volume fractions, depending on the injection-molding layout and material design. Therefore, it is infeasible to reconstruct a new SFRC RVE for each individual microstructural geometry. To reduce the cost of DMN training, we introduce a transfer learning scheme *(Liu et al. 2019b; Liu et al. 2020; Huang et al. 2022)* to quickly generate DMNs for new SFRC microstructures by transferring the knowledge of a few pre-trained networks. To consider the effects of different fiber orientation states, we need to reconstruct 3 SFRC RVE geometries with the same fiber volume fraction, including RVEs with random 3D, random 2D, and unidirectional (UD) fiber orientation states. These three special orientation states are chosen for the offline training because a linear combination of their corresponding second-order orientation tensors *(Advani and Tucker III, 1987)* is sufficient for parametrization of all other possible fiber orientation states, as explained in **Appendix II** of this paper. Furthermore, an additional RVE geometry with a UD fiber orientation state and a high fiber volume fraction is reconstructed to capture the fiber volume fraction effect on the composite response. In total, we have reconstructed 4 SFRC RVE geometries with a fiber aspect ratio around 20 for the offline training of DMN, as shown in Fig. 2.

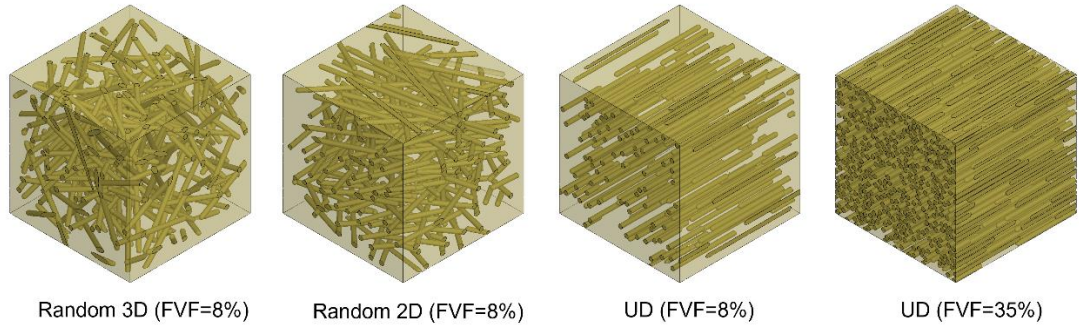

Random 3D (FVF=8%)    Random 2D (FVF=8%)    UD (FVF=8%)    UD (FVF=35%)

**Fig. 2.** Short-fiber-reinforced composite microstructures for transfer-learning-based offline training of DMN models, where FVF denotes the fiber volume fraction.

For each SFRC microstructural geometry, we define 500 material samples containing different microscopic stiffness tensors for the fiber phase and matrix phase, and computational homogenization is employed to obtain the corresponding macroscopic stiffness tensors for the composites. Material samples are assigned with linearly elastic microscopic stiffness tensors with sufficient phase contrast and material anisotropy *(Liu and Wu 2019)* for the material network to learn the topological representation of SFRC. In the present study, each microstructure is discretized by 10-node tetrahedron finite elements in LS-DYNA and the *RVE_ANALYSIS_FEM keyword is used to automatically impose the periodic displacement boundary conditions for homogenization (*Wei and Lyu et al. 2022*). Since 4 microstructural geometries are considered for offline training, 2000 linear elastic finite element models are generated in total. To calculate the macroscopic composite stiffness tensor, 6 orthogonal loading conditions are imposed to each finite element model, respectively. As a result, 12000 linear elastic finite element simulations are performed, for which the total CPU time is approximately 670 hours with 32 processors used for each simulation.

After the finite element simulations, the homogenized composite stiffness and the corresponding microscopic fiber and matrix stiffness data are collected, and then the material data for each RVE microstructural geometry are separated into two datasets, of which 400 data points are defined as the training dataset, and the remaining 100 data points are defined as the testing dataset. The training dataset is utilized with a gradient-based optimization to calculate the network parameters of DMN models during the offline training stage, whereas the testing dataset is used to assess the generalization performance of a trained model.

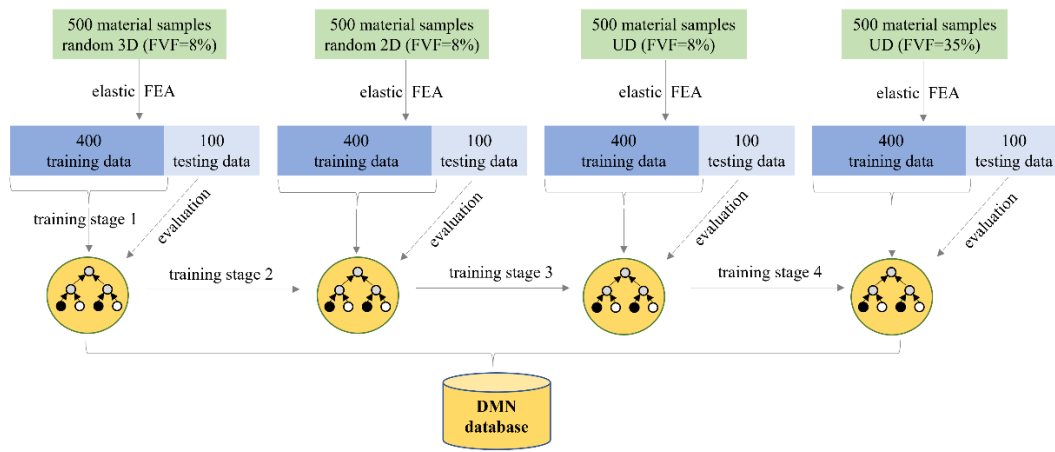

**Fig. 3.** Workflow for the 4-stage offline training of DMN for SFRC.

Transfer learning-based offline training of DMN for SFRC consists of the following four stages, as illustrated in Fig. 3. In stage 1, the RVE microstructure with 8% fibers uniformed oriented in 3D is considered, and DMN is trained with randomly initialized trainable parameters. After this stage, we obtain a trained DMN model, and we transfer it to initialize the networks for the random 2D and the UD RVEs with 8% fibers in stage 2 and stage 3, respectively. Finally, in stage 4 we transfer the network trained for the UD RVE with a low fiber volume fraction at 8% to the UD RVE with a high fiber volume fraction at 35%. For each stage, we use 20000 epochs to train a DMN model, where one epoch refers to one round of evaluation on all the training samples. During the optimization process, the 400 training samples for each SFRC microstructure are divided randomly into 10 mini-batches, so there are 10 training steps in each epoch. In addition, the bold driver method is employed to adapt the learning rate (i.e., the multiplier on the step size) by comparing the training error to its previous value after each epoch. Parallel computing with 10 processors is adopted for the offline training, and it takes around 200 hours to finish all the 4 training stages.

Histories of the average training and testing errors for the offline training process are plotted in Fig. 4, where the average errors are defined in (*Liu and Wu 2019*). DMN for the first RVE with the random 3D microstructure begins with a large training error since the trainable parameters are randomly initialized without any prior knowledge about the microstructure. For the other three RVEs, the training starts from a much lower error thanks to the knowledge transferred from the pre-trained network, which demonstrates the enhanced training efficiency of the employed transfer learning scheme.

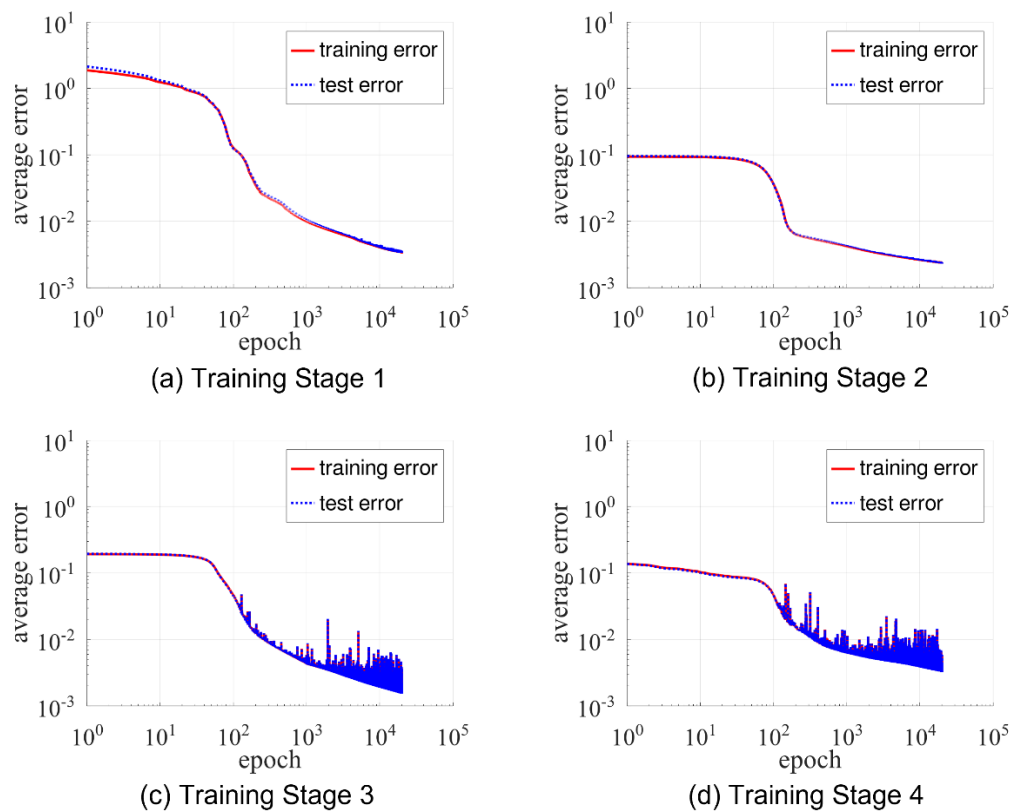

**Fig. 4.** Histories of the average training and testing errors for transfer-learning-based training of DMN models for SFRC.

**Table 1.** Training results of DMN for SFRC microstructures

| Microstructures | Random 3D FVF=8% | Random 2D FVF=8% | UD FVF=8% | UD FVF=35% |
|---|---|---|---|---|
| Training error | 0.33% | 0.23% | 0.16% | 0.31% |
| Testing error | 0.33% | 0.23% | 0.15% | 0.30% |

Table 1 shows the training results for DMN with 8 layers, where the accuracy is measured by the scaled mean absolute error. As can be seen from the table, training errors of all the DMN models are less than or equal to 0.33%. In addition, we can observe that the training error decreases from the random 3D fiber orientation to random 2D fiber orientation, and it further decreases for the UD fiber orientation state. Increasing the fiber volume fraction, however, induces a higher training error. The levels of testing errors on unseen data points are quite close to the training error levels, suggesting that there is no overfitting issue. The strong generalization performance of DMN is attributed to the essential physics embedded in the two-layer building block, which enhances the extrapolation capability to unknown material and loading spaces. To further examine the ability of trained DMN for capturing material anisotropic effects, DNS and DMN are employed to predict the homogenized linear elastic material stiffness for UD and random 2D SFRC microstructures, respectively, for which we adopt a set of fiber and matrix properties unseen in the training process. Using the method described by (*Nordmann et al. 2018*), direction-dependent Young's modulus calculated from the anisotropic stiffness can be visualized as a 3D surface, as shown in Fig. 5. The good agreement between DNS and DMN predictions confirms the effectiveness of DMN for capturing the microstructure-induced directional dependency of SFRC properties.

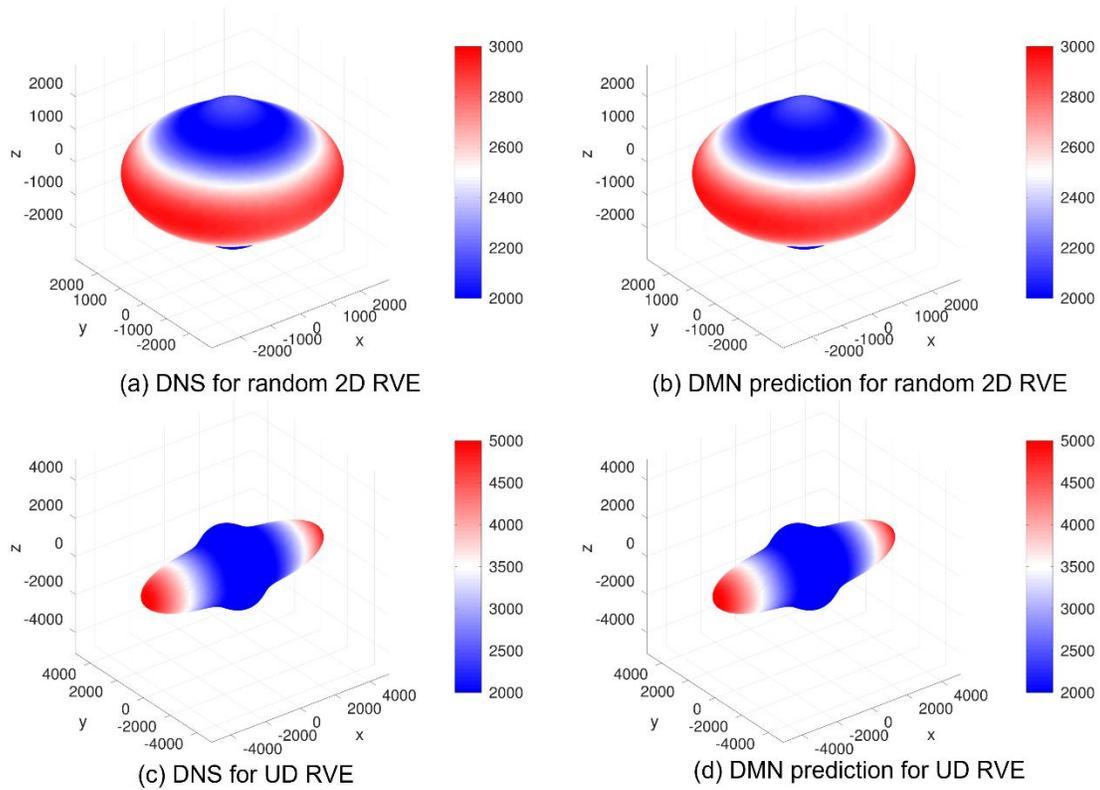

**Fig. 5.** 3D representation of Young's modulus for anisotropic SFRC microstructures predicted by DNS and DMN, where the radius (vector measured from the origin to the surface) in any direction is proportional to the magnitude of the Young's modulus in that direction, and the magnitude of the Young's modulus is also conveyed by a color mapping applied to the surface.

### LS-DYNA Nonlinear Multiscale Online Prediction for Injection-Molded SFRC

Integration of DMN models for SFRC with the engineering simulation software LS-DYNA is implemented for multiscale structural analysis of SFRC. In dynamic finite element analysis, the spatial domain $V$ of the global SFRC structure is discretized into a collection of subdomains $V_e$, where $e = 1, \cdots, N_e$, $N_e$ denotes the total number of finite elements, and the global displacement field $\boldsymbol{u}(\boldsymbol{X}, t) \in \mathbb{R}^3$ is approximated by

$$\boldsymbol{u}(\boldsymbol{X}, t) = \sum_{I=1}^{N_n} \mathrm{N}_I^u(\boldsymbol{X}) \mathbf{U}_I(t) \tag{10}$$

where $I$ denotes the global nodal index, $N_n$ denotes the total number of nodes in the finite element mesh, $\boldsymbol{X} \in \mathbb{R}^3$ and $t$ denote the spatial position and time, respectively, $\mathrm{N}_I^u(\boldsymbol{X})$ and $\mathbf{U}_I(t) \in \mathbb{R}^3$ denote the shape function and the displacement vector associated with node $I$, respectively. The nodal displacements, velocities, and accelerations can be obtained by solving the global semi-discrete momentum equation:

$$\mathbf{M}\ddot{\mathbf{U}} = \mathbf{F}^{\text{ext}} - \mathbf{F}^{\text{int}} \tag{11}$$

where $\ddot{\mathbf{U}} \in \mathbb{R}^{3N_n}$ is the global nodal acceleration vector, which is the 2nd-order material time derivative of the global nodal displacement vector $\mathbf{U} \in \mathbb{R}^{3N_n}$, $\mathbf{M} \in \mathbb{R}^{(3N_n) \times (3N_n)}$, $\mathbf{F}^{\text{ext}} \in \mathbb{R}^{3N_n}$, and $\mathbf{F}^{\text{int}} \in \mathbb{R}^{3N_n}$ are the lumped mass matrix, the global external nodal force vector, and the global internal nodal force vector, respectively. $\mathbf{F}^{\text{int}}$ is obtained by assembling all the internal force vectors $\mathbf{F}_I^{\text{int}}$ associated with every node defined by

$$\mathbf{F}_I^{\text{int}} = \int_V \boldsymbol{\mathcal{B}}_I^T \, \boldsymbol{\sigma} \, \mathrm{d}V \tag{12}$$

where $\boldsymbol{\mathcal{B}}_I$ denotes the shape function gradient matrix associated with node $I$. Evaluation of the internal nodal force in Eq.(12) is based on numerical integration. To this end, the macroscopic stress $\boldsymbol{\sigma}$ is calculated at every quadrature point $\boldsymbol{\xi}_q$ of the finite element model, where the subscript $q$ denotes the quadrature point index. In the present LS-DYNA multiscale method, the macroscopic stress $\boldsymbol{\sigma}$ is predicted by DMN coupled with finite elements, where each quadrature point $\boldsymbol{\xi}_q$ has an associated network corresponding to the local fiber orientation and volume fraction. To model injection-molded SFRC parts, where nonuniform fiber distributions are induced by different molding conditions (e.g., part geometry, injection gate position, filling time, and mold temperature), it is desirable to create the online DMN models in an efficient manner, instead of performing offline training for each individual microstructure. For this reason, the transfer learning method proposed in (*Liu et al. 2019b; Liu et al. 2020; Huang et al. 2022*) is adopted for creating DMN models in LS-DYNA during online computation. Under the transfer learning framework, the base topological structures of all the four DMN models obtained from offline training are analogous, which enables a continuous migration between different networks through direct interpolation of their trainable parameters. Let us define a data point $(X^*, Y^*)$, where the superscript $(*)$ denotes an intermediate state, $Y^*$ denotes the unknown DMN trainable parameters:

$$Y^* = [z^*, \alpha^*, \beta^*, \gamma^*] \tag{13}$$

and $X^*$ denotes the geometric descriptors of the intermediate SFRC microstructure:

$$X^* = [v_f^*, a_{11}^*, a_{22}^*] \tag{14}$$

in which $v_f^*$ denotes the fiber volume fraction; $a_{11}^*$ and $a_{22}^*$ are two largest eigenvalues of the second-order fiber orientation tensor (*Advani and Tucker III, 1987*), which describes the orientation state of short fibers. Note that the three eigenvalues $a_{11}$, $a_{22}$, and $a_{33}$ of any fiber orientation tensor $a$ satisfy $a_{11} \geq a_{22} \geq a_{33}$ and $a_{11} + a_{22} + a_{33} = 1$, as described in **Appendix II**. The values of fiber orientation tensor and volume fraction can be either measured from experiments or predicted through injection molding simulation of the melt flow process (*Wang et al. 2018*). Similarly, we can define the known trainable parameters of pre-trained DMN models as $Y_1$, $Y_2$, …, $Y_N$, and the geometric descriptors of microstructures used in the offline training as $X_1$, $X_2$, …, $X_N$. Accordingly, the regression function for the new data point $(X^*, Y^*)$ can be expressed as

$$Y^*(X^*) = r(X^* | (X_1, Y_1), (X_2, Y_2), \cdots, (X_N, Y_N)) \tag{15}$$

To determine the unknown trainable parameters (i.e., $[z^*, \alpha^*, \beta^*, \gamma^*]$) for a linear regression model with three independent geometric descriptors (i.e., $[v_f^*, a_{11}^*, a_{22}^*]$), we need four linearly independent data points $(X_i, Y_i)$, which correspond to the four RVE geometries created in the offline training stage. Therefore, N=4 is chosen in Eq. (15).

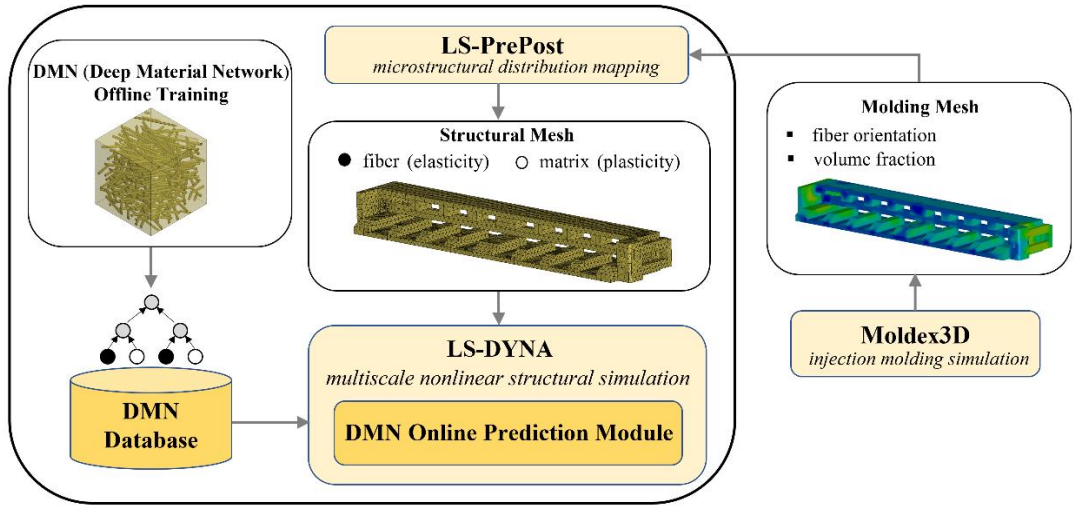

**Fig. 6.** Illustration of the DMN-based nonlinear multiscale simulation framework for short-fiber-reinforced composite (SFRC) structures, where microstructural data from Moldex3D are mapped by LS-PrePost to LS-DYNA finite elements coupled with DMN.

Since the online network creation is guided by the microstructures at quadrature points, it is essential to gather the injection-molded microstructure information. In practice, microstructural distribution in SFRC products can be obtained through injection molding simulation (*Wang et al. 2018*) using the software Moldex3D, and the predicted fiber orientation and volume fraction data can be mapped from the molding simulation mesh to the LS-DYNA structural simulation mesh using the pre-processing software LS-PrePost. After mapping, the DMN online prediction module will create a new DMN model at each quadrature point specific to the local microstructure, and then the network will be dynamically coupled to the finite elements in LS-DYNA for nonlinear multiscale online prediction. An illustration of the overall multiscale simulation framework (*Wei et al. 2021*) is depicted in Fig. 6. Note that this online DMN creation process does not involve RVE reconstruction or DNS. In addition, the new DMN models are created only once at the beginning of the online prediction stage, so the associated computational cost is negligible in the overall multiscale simulation.

After the creation of microstructure-based DMN models, LS-DYNA multiscale structural simulations will be carried out, where finite element modeling for the global structures and DMN prediction of the local composite materials are tightly coupled. At each time step, finite element equations are solved to calculate the nodal accelerations, velocities, and displacements at the global structural level. In the present work, an explicit time integration algorithm has been adopted, which has been proven to be highly efficient and robust for nonlinear dynamic problems involving contact-impact and large deformations (*Belytschko et al. 2014*). Afterwards, the macroscopic strain at each quadrature point is evaluated and transferred to DMN. With the macroscopic strain increment, backward de-homogenization and forward homogenization of material information are performed within DMN to predict the multiscale material response. The incremental stress-strain relationship associated with DMN's $k^{\text{th}}$ node at layer $i$ takes the following form:

$$\Delta \overline{\boldsymbol{\sigma}}_i^k = \overline{\boldsymbol{C}}_i^k \Delta \overline{\boldsymbol{\varepsilon}}_i^k + \mathrm{d}\overline{\boldsymbol{\sigma}}_i^k \tag{16}$$

Here, $\overline{\boldsymbol{C}}_i^k$ is the averaged material stiffness, $\Delta \overline{\boldsymbol{\varepsilon}}_i^k$ denotes the strain increment of DMN's $k^{\text{th}}$ node at layer $i$. In multiscale structural analysis, macroscopic rate-of-deformation increments computed by the finite element method are assigned to the top layer node of DMN at the corresponding quadrature point. Strain increments of nodes at other layers are calculated through backward de-homogenization. $\mathrm{d}\overline{\boldsymbol{\sigma}}_i^k$ denotes a correction to the incremental stress, which should vanish if material nonlinearities of composites are omitted. In nonlinear composite modeling, however, $\mathrm{d}\overline{\boldsymbol{\sigma}}_i^k$ is not necessarily equal to zero and is calculated through forward propagation from a lower layer of the network:

$$\mathrm{d}\overline{\boldsymbol{\sigma}}_i^k = \frac{w_{i+1}^{2k-1}}{w_{i+1}^{2k-1}+w_{i+1}^{2k}}\mathrm{d}\boldsymbol{\sigma}_{i+1}^{2k-1} + \frac{w_{i+1}^{2k}}{w_{i+1}^{2k-1}+w_{i+1}^{2k}}\mathrm{d}\boldsymbol{\sigma}_{i+1}^{2k} + \boldsymbol{\chi} \tag{17}$$

where $w_{i+1}^{2k-1}$ and $w_{i+1}^{2k}$ are the corresponding nodal weights, and the vector $\boldsymbol{\chi}$ depends on the material stiffness matrices and stress corrections of the two child nodes, for which an analytical expression can be found in (*Liu and Wu, 2019*). $\mathrm{d}\boldsymbol{\sigma}_{i+1}^{2k-1}$ and $\mathrm{d}\boldsymbol{\sigma}_{i+1}^{2k}$ are the rotated stress corrections of child nodes, which are obtained by applying a rotation operation to the averaged stress correction:

$$\mathrm{d}\boldsymbol{\sigma}_i^j = \boldsymbol{R}\big(\alpha_i^j,\beta_i^j,\gamma_i^j\big)\,\mathrm{d}\overline{\boldsymbol{\sigma}}_i^j \tag{18}$$

where $\boldsymbol{R}\big(\alpha_i^j,\beta_i^j,\gamma_i^j\big)$ denotes the rotation matrix based on the Euler angles $\big(\alpha_i^j,\beta_i^j,\gamma_i^j\big)$ of the network. At the bottom layer of DMN, material stiffness matrices $\boldsymbol{C}_N^k$, incremental stress $\Delta \boldsymbol{\sigma}_N^k$, and the correction $\mathrm{d}\boldsymbol{\sigma}_N^k$ are evaluated using microscopic constitutive laws for the fiber phase and the matrix phase. While linear elastic constitutive laws are adopted during the offline stage to learn the essential physics, elastoplastic microscopic constitutive laws can be adopted in the online structural analysis stage to capture nonlinear composite material behaviors. For SFRC, a linear elastic law is usually sufficient for modeling the fiber phase, whereas an elastoplastic law with isotropic hardening can be adopted for modeling the nonlinear matrix phase. After the microscopic material law evaluation, stress and state variables (e.g., equivalent plastic strain/EPS) are stored at the bottom layer, while the stiffness matrices and stress corrections are propagated to an upper layer of the network. Due to material nonlinearities, forward homogenization and backward de-homogenization of stresses and strains are iterated in the network. To check convergence for the network iteration, an $L_2$ norm of the difference in two successive strains is computed at the bottom layer:

$$\sum_{k=1}^{2^{N-1}} \left\| \Delta \boldsymbol{\varepsilon}_N^{k\,(ite+1)} - \Delta \boldsymbol{\varepsilon}_N^{k\,(ite)} \right\| \leq \epsilon_{tol} \tag{19}$$

where the superscripts $(ite)$ and $(ite+1)$ denote iteration counts, and $\epsilon_{tol}$ is a convergence tolerance. Once convergence is achieved, the microscopic stress and state variables (e.g., equivalent plastic strain) of each bottom node are updated, and the stress increment $\Delta \boldsymbol{\sigma}_1^1$ of the DMN's top layer node is employed to update the macroscopic stress $\boldsymbol{\sigma}$ at the finite element's quadrature point $\boldsymbol{\xi}_q$ :

$$\boldsymbol{\sigma}\big(\boldsymbol{\xi}_q\big)^{t_{n+1}} = \boldsymbol{\sigma}\big(\boldsymbol{\xi}_q\big)^{t_n} + \Delta \boldsymbol{\sigma}_1^1\big(\boldsymbol{\xi}_q\big) \tag{20}$$

where the superscript $t_n$ and $t_{n+1}$ denote two different time instants during the time integration of the momentum equation. Upon the completion of the DMN-based multiscale stress computation, finite elements in LS-DYNA will gather the macroscopic stress from different quadrature points to evaluate the internal force vector $\mathbf{F}_I^{\text{int}}$ by Eq.(12) for nonlinear finite element analysis. After the internal force computation, the resulting finite element equations for composite structures can be solved for the next time step. A flowchart for the DMN-based internal force calculation in LS-DYNA is given in Box 1. It is noteworthy to mention that, in addition to applying DMN in the nonlinear finite element modeling, it is also feasible to couple DMN with meshfree methods (*Wang et al. 2009; Wu et al. 2020; Huang et al. 2020; Pasetto et al. 2021*) for accelerated multiscale analysis of structures undergoing extreme deformations.

---

**Box 1. Flowchart for DMN-based internal force calculation in FEA**

a.    Initialization: $\mathbf{F}^{\text{int}} = \mathbf{0} \in \mathbb{R}^{3N_n}$

b.    Loop over finite elements $e = 1, \cdots, N_e$

     **i.**    Gather element nodal displacements and velocities

     **ii.**    Loop over quadrature points $\boldsymbol{\xi}_q \in \mathbb{R}^3$ with quadrature weights $\varpi_q(\boldsymbol{\xi}_q)$

         **1.**    Initialize Deep Material Network (DMN) parameters if time $\mathrm{t}_n = 0$

           1.1    Import fiber orientation $\boldsymbol{a}(\boldsymbol{\xi}_q) \in \mathbb{R}^{3\times3}$, volume fraction $v_f(\boldsymbol{\xi}_q)$

           1.2    Regression-based transfer learning to get new network parameters $\boldsymbol{z}$, $\boldsymbol{\alpha}$, $\boldsymbol{\beta}$, $\boldsymbol{\gamma}$, $\boldsymbol{w}$ based on SFRC microstructure at point $\boldsymbol{\xi}_q$

           Retrieve DMN parameters $\boldsymbol{\alpha}$, $\boldsymbol{\beta}$, $\boldsymbol{\gamma}$, $\boldsymbol{w}$ stored at point $\boldsymbol{\xi}_q$ if time $\mathrm{t}_n > 0$

         **2.**    Compute macroscopic rate-of-deformation increment $\Delta \mathbf{D}(\boldsymbol{\xi}_q)$

         **3.**    Compute Cauchy stress increment $\Delta\boldsymbol{\sigma}(\boldsymbol{\xi}_q)$ by DMN

           3.1    Evaluate microscopic constitutive equations to get stress $\Delta\bar{\boldsymbol{\sigma}}_N^k$, $\mathrm{d}\bar{\boldsymbol{\sigma}}_N^k$ stiffness $\overline{\boldsymbol{C}}_N^k$, and material state variables of bottom-layer nodes

           3.2    Forward homogenization of stress $\mathrm{d}\bar{\boldsymbol{\sigma}}_i^k$ and stiffness $\overline{\boldsymbol{C}}_i^k$

           3.3    Compute stress increment at the top layer $\Delta\boldsymbol{\sigma}_1^1 = \boldsymbol{C}_1^1 \cdot \Delta\mathbf{D}(\boldsymbol{\xi}_q) + \mathrm{d}\boldsymbol{\sigma}_1^1$

           3.4    Backward de-homogenization of stress $\Delta\bar{\boldsymbol{\sigma}}_i^k$ and strain $\Delta\bar{\boldsymbol{\varepsilon}}_i^k$

           3.5    Check network convergence. If not converged, go to 3.1

         **4.**    Update macroscopic Cauchy stress $\boldsymbol{\sigma}(\boldsymbol{\xi}_q) \leftarrow \boldsymbol{\sigma}(\boldsymbol{\xi}_q) + \Delta\boldsymbol{\sigma}_1^1$

         **5.**    Update internal nodal force, $\mathbf{F}_I^{\text{int}} \leftarrow \mathbf{F}_I^{\text{int}} + \boldsymbol{\mathcal{B}}_I^T(\boldsymbol{\xi}_q)\boldsymbol{\sigma}(\boldsymbol{\xi}_q)\varpi_q(\boldsymbol{\xi}_q)$

     **iii.**    Assemble $\mathbf{F}_I^{\text{int}}$ to global internal nodal force vector $\mathbf{F}^{\text{int}}$

     **iv.**    END loop over quadrature points

c.    END loop over finite elements

## Applications for Nonlinear Modeling of Short-Fiber-Reinforced Composites

In this section, two numerical examples are presented to demonstrate the effectiveness and performance of the present DMN-based multiscale method. In the first example, we verify the accuracy and efficiency of the method by comparing with direct numerical simulations of RVE, where both the microstructural geometries and nonlinear microscopic material laws are unseen in the DMN offline training. In the second example, nonlinear multiscale analysis is performed for a short-fiber-reinforced thermoplastic part by integrating injection molding-induced fiber orientations and volume fractions, which demonstrates the capability of the present method for industrial applications where capturing the microstructural effects is essential.

### *Verification Against Direct Numerical Simulation of SFRC RVE*

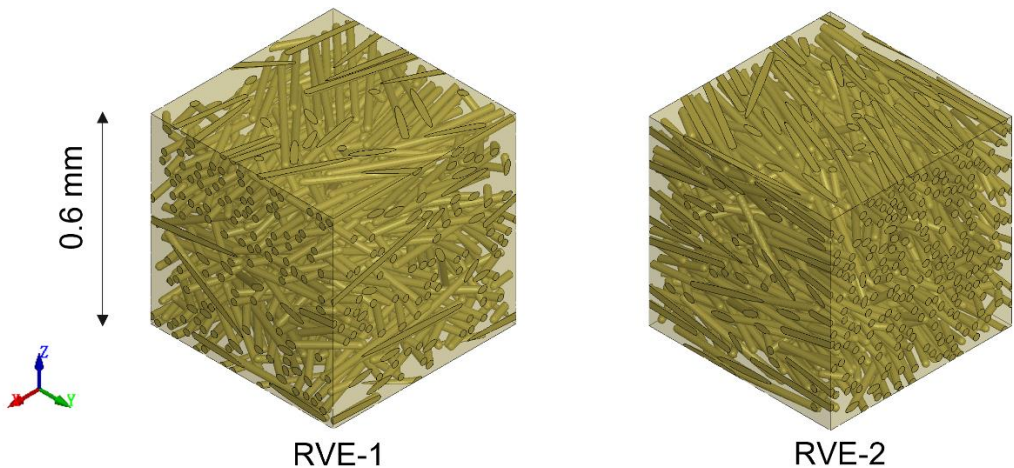

**Fig. 7.** Reconstructed SFRC microstructures for direct numerical simulation.

**Table 2.** SFRC microstructures analyzed in the online prediction

| SFRC RVE | Fiber Orientation Tensor | | | | | | Fiber Volume Fraction |
| --- | --- | --- | --- | --- | --- | --- | --- |
| | $a_{xx}$ | $a_{yy}$ | $a_{zz}$ | $a_{xy}$ | $a_{yz}$ | $a_{zx}$ | |
| 1 | 0.5861 | 0.3521 | 0.0618 | 0.05447 | -0.0172 | -0.0159 | 19.4% |
| 2 | 0.1353 | 0.8036 | 0.0611 | 0.1504 | -0.009521 | -0.005788 | 24.0% |

In this example, we present nonlinear online prediction results of DMN at a single macroscopic material point level for different SFRC microstructures. The two analyzed

SFRC microstructures are illustrated in Fig. 7, and Table 2 provides the fiber orientations and volume fractions based on typical microstructures observed in injection-molded SFRC parts. For thermoplastics, a nonlinear isotropic elastoplastic material model with a piecewise linear hardening law is employed, whereas for glass fibers an isotropic linear elastic material model is adopted, and the material properties are given in Table 3. The microstructures and the nonlinear microscopic material models are unseen in the DMN offline training stage.

**Table 3.** Material properties of SFRC constituents for the RVE simulation

|  | Matrix phase | Fiber phase |
| --- | --- | --- |
| Young's modulus | 1616 MPa | 72000 MPa |
| Poisson ratio | 0.3545 | 0.20 |
| Initial tensile yield strength | 0.63 MPa | --- |
| Mass density | $1.0 \times 10^{-9}$ tonne/mm$^3$ | $2.54 \times 10^{-9}$ tonne/mm$^3$ |

**Table 4.** High-fidelity finite element discretization adopted in DNS of SFRC RVE

|  | RVE 1 | RVE 2 |
| --- | --- | --- |
| number of elements | 4450825 | 5047295 |
| number of DOF | 18226953 | 20622111 |

As a comparison, we conducted direct numerical simulations (DNS) of RVE models in LS-Dyna. The RVE size is set to be about 1.3 times the average value of the fiber length, and 10-node tetrahedron finite elements are adopted to achieve a high-fidelity discretization, for which the number of elements and the number of displacement degrees of freedom (DOF) are summarized in Table 4. An unconstrained uniaxial tensile loading condition is imposed by applying a periodic displacement boundary condition on the finite element model. Nonlinear implicit computation is conducted, and macroscopic stress-strain results are obtained through computational homogenization. As shown in Fig. 8 and Fig. 9, localized plasticity in the matrix phase and concentrated stress in the fiber phase appear in the composites, leading to highly nonlinear elastoplastic behaviors. In addition, different microstructural geometries lead to distinct responses of the composites.

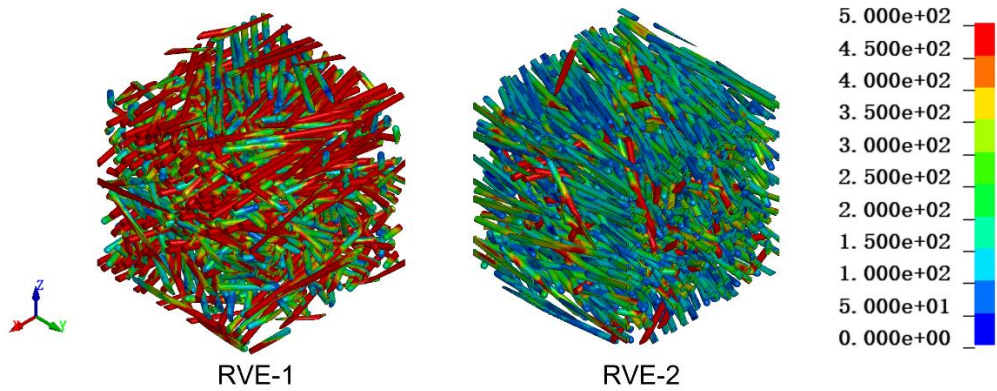

**Fig. 8.** DNS predicted von Mises stress in the fiber phase of SFRC (plotted on the undeformed RVE configuration).

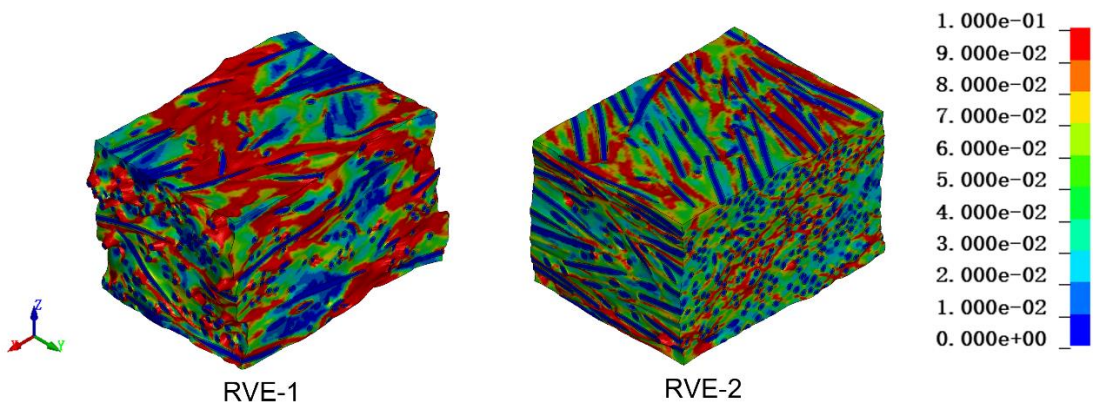

**Fig. 9.** DNS predicted equivalent plastic strain in the matrix phase of SFRC (plotted on the deformed RVE configuration with a deformation scale factor 5.0).

For DMN-based online prediction, a single solid finite element is coupled with DMN, whose network parameters are generated through transfer learning considering the fiber orientation tensor and volume fraction of RVE-1 and RVE-2, respectively. After the new DMN network is formed in the online stage, the macroscopic strain tensor predicted by DNS is enforced at the top node of the network, which ensures a consistent macroscopic tensile deformation state for the DNS and the corresponding DMN simulations.

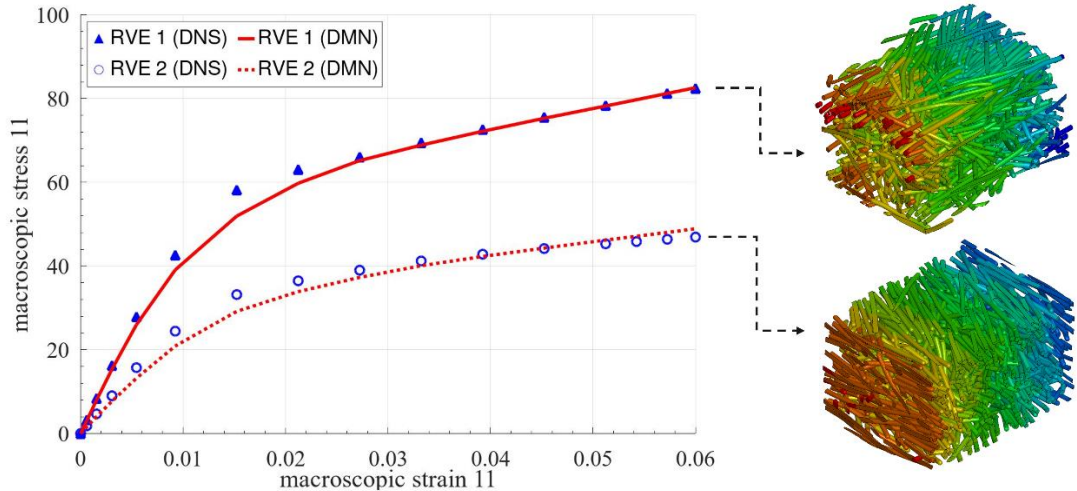

**Fig. 10.** Macroscopic stress-strain curves of SFRC RVE predicted by DMN and DNS with corresponding displacement fields of the deformed RVE (scale factor 5.0 is used to show the deformed configuration).

The macroscopic stress-strain curves predicted by DMN and DNS are plotted in Fig. 10. As can be seen, DMN well captures the influence of complex SFRC microstructures on the overall elastoplastic material behaviors of composites. Specifically, RVE-1 shows a stiffer mechanical response over RVE-2 along the tensile loading direction, which is naturally predicted by the microstructure-sensitive DMN. Overall speaking, a satisfactory agreement is achieved between the DMN-based nonlinear prediction results and the FEM-based high-fidelity DNS results. It is worthwhile to mention that DNS is often infeasible for SFRC RVE due to the challenges in mesh generation for RVE with high fiber volume fractions, and occasionally the finite element simulation may experience numerical convergence issues due to strong nonlinearity and mesh distortions. All these issues are naturally circumvented by the proposed DMN-based multiscale modeling approach.

**Table 5.** Total CPU time of DMN and DNS for SFRC RVE modeling

|  | RVE 1 | RVE 2 |
|---|---|---|
| DNS (64 cores) | 32.25 hours | 40.38 hours |
| DMN (1 core) | 1 second | 1 second |

Computational costs of DMN and DNS for the SFRC RVE modeling are summarized in Table 5. We can see that the prediction using DMN on 1 core is 100~150 thousand times faster than the finite element-based DNS on 64 cores. The computational burden of DNS is mainly related to solving the system of finite element equations with approximately 20 million degrees of freedom per RVE model, which consumes high CPU time and memory despite the usage of a parallel iterative equation solver. In addition, the finite element model for each RVE contains approximately 20 million integration points where material laws must be evaluated at each time step, which further slows down the computation. In contrast, the computational cost of DMN is mainly associated with the material law evaluation at the bottom layer nodes. For an 8-layer network, there are only 128 bottom nodes, thus the computational speed is several orders-of-magnitude faster than FEM-based DNS. In DNS, nonlinear deformations of RVE models occasionally lead to distorted finite element shapes, which results in convergence difficulties during the implicit analysis. In this scenario, decreased load step sizes and increased numbers of iterations must be adopted in DNS. On the other hand, satisfactory convergence is achieved in DMN simulations since mesh entanglements are naturally avoided in the network iteration algorithm.

Clearly, DMN can be seen as an effective and robust reduced-order model of the high-fidelity DNS model. As described in the precious section, there are certain computational costs during the offline training stage for creation of linear elastic training data and optimization of network parameters. Nevertheless, once the offline training is finished, the trained DMN models can be integrated with FEM for online prediction to significantly accelerate the nonlinear multiscale simulation.

### *Nonlinear Multiscale Simulation of An Injection-Molded Car Component*

This example presents an industrial application of the present LS-DYNA machine learning-based multiscale method for nonlinear dynamic simulation of SFRC parts. As shown in Fig. 11, we model a dynamic impact-contact process involving an automotive part crashing into a rigid pole with an initial velocity of 40 m/s, for which a finite element mesh consisting of 175509 nodes and 682452 elements (682452 tetrahedron solids for the SFRC part and 2200 shells for the rigid pole) is generated.

To initiate DMN models in LS-DYNA multiscale structural simulation, it is essential to obtain the injection-molding-induced microstructure distribution in the SFRC part. To this end, injection molding simulation of the filling, packing, and cooling processes is performed using the Moldex3D software. Fig. 12 shows the injection molding set-up and the numerical model for molding simulation, which contains 1058868 finite volume elements and 472501 nodes. Typically, the molding mesh is not identical to the structural mesh, as different factors need to be considered for the discretization of fluids and solids, respectively. For instance, the molding mesh requires a locally refined discretization near the injection gate location to accurately capture the melt inflow, as seen in Fig. 12(b). Fig. 13 plots the predicted fiber orientation tensor distribution, which shows a strong location dependency of the fiber alignment. In addition, although the predicted fiber volume fraction is around 18%, nonuniform fiber concentrations can be clearly seen near the two injection gates. The heterogeneous microstructural distribution is then mapped from the molding mesh shown in Fig. 12(b) onto the structural mesh shown in Fig. 11(b) using the pre-processing software LS-PrePost. The mapped data serve to guide the creation of new DMN models specific to the local fiber orientation and volume fraction at every quadrature point of the structural mesh, following the transfer learning strategy.

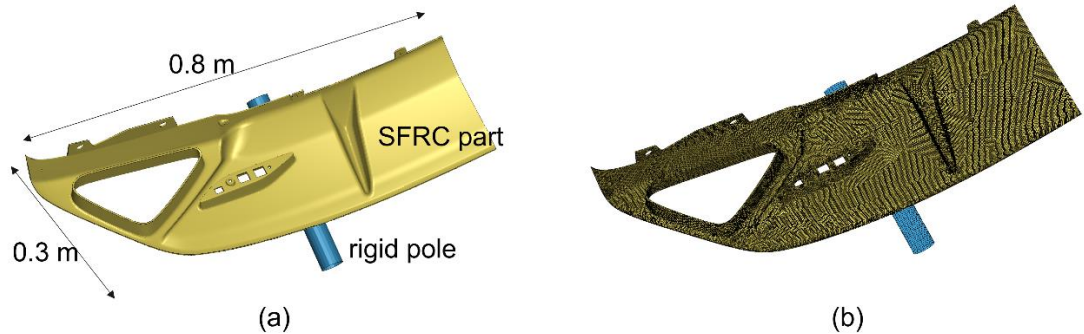

**Fig. 11.** An automotive part made of injection-molded short-fiber-reinforced thermoplastic composites. (a) Geometry of the SFRC part and the rigid pole. (b) Solid finite element mesh used in the nonlinear multiscale structural simulation.

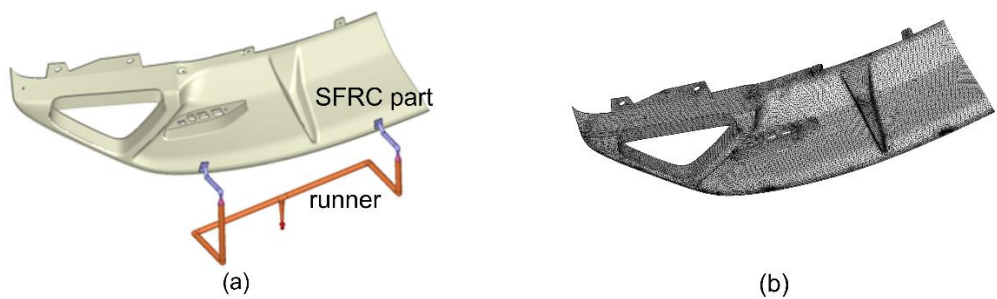

**Fig. 12.** An automotive part made of injection-molded short-fiber-reinforced thermoplastic composites. (a) Geometry of the SFRC part and the hot runner for injection molding. (b) Solid finite volume mesh used in the injection molding simulation.

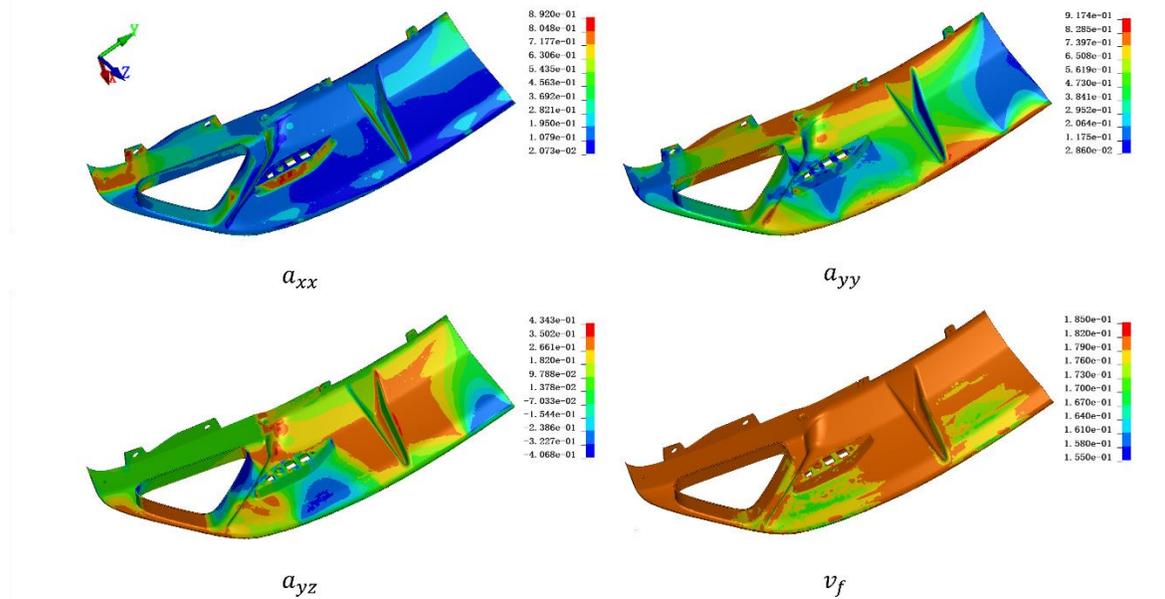

**Fig. 13.** Manufacturing process-induced microstructural distribution predicted by the injection molding simulation, where fiber orientation tensor components $a_{xx}$, $a_{yy}$, $a_{yz}$ and the fiber volume fraction $v_f$ are plotted.

**Table 6.** Material properties of SFRC constituents for the automotive part simulation

|  | Matrix phase | Fiber phase |
|---|---|---|
| Young's modulus | 3800 MPa | 80000 MPa |
| Poisson ratio | 0.39 | 0.20 |

In the multiscale structural analysis, we employ an elastic model for fibers and a nonlinear plasticity model for the matrix phase, for which the elastic constants are given in Table 6. To describe the elastoplastic behavior of the matrix, we adopt the following von Mises yield function with isotropic hardening:

$$s_Y^m = s_1^m + s_2^m \cdot \bar{\varepsilon}_P^m - s_3^m \cdot \exp(-h_0^m \cdot \bar{\varepsilon}_P^m)$$

where $s_Y^m$ denotes the yield strength for the matrix phase, $\bar{\varepsilon}_P^m$ denotes the accumulated equivalent plastic strain of the matrix material, and plastic yielding parameters $h_0^m = 140.0$, $s_1^m = 120.0$ MPa, $s_2^m = 0.0$ MPa, and $s_3^m = 90.0$ MPa are chosen. The SFRC part moves toward the rigid pole with an initial velocity of 40 m/s, and the total simulation time is 1.8 ms.

During the simulation, the contact force induced by the dynamic interaction between the composite part and the rigid pole is measured. Time history of the resultant contact force is plotted in Fig. 14, where homogenized von Mises stress on the deformed SFRC part is also plotted at three different time instants, and Fig. 15 plots the evolution of homogenized equivalent plastic strain (EPS) due to plastic deformations in the matrix phase. In these figures, the von Mises stress is calculated from the homogenized stress tensor at DMN's top layer, whereas the homogenized EPS is obtained by forward propagation of microscopic EPS in the matrix phase at DMN's bottom layer. The results clearly demonstrate the capability of the present multiscale method for capturing complicated dynamic plastic deformations of short-fiber-reinforced composite structures.

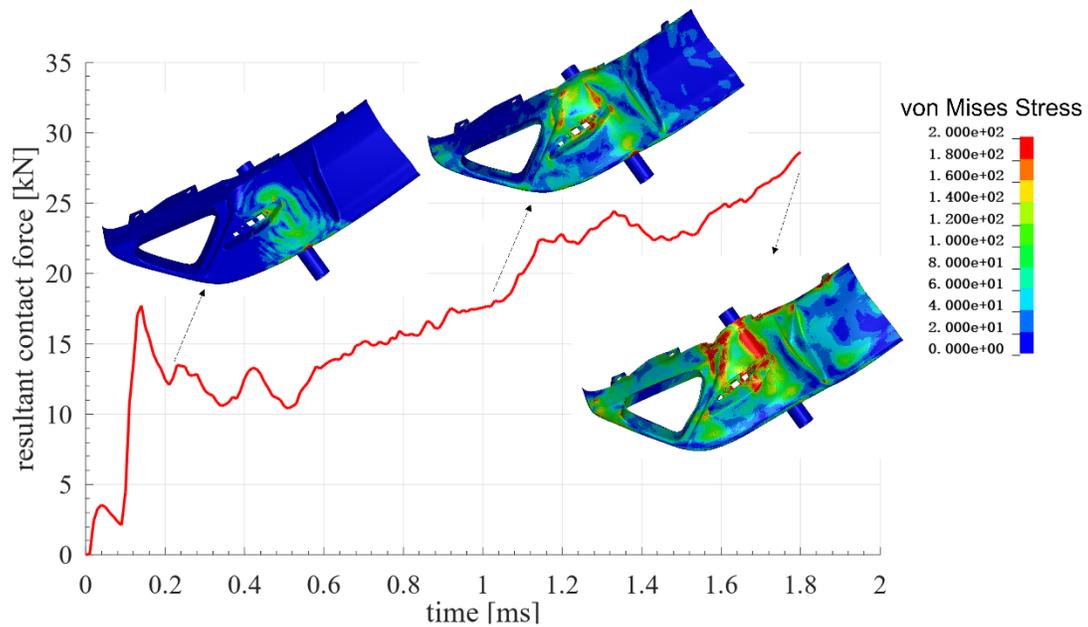

**Fig. 14.** Time history of the resultant contact force, where homogenized von Mises stress distribution on the deformed short-fiber-reinforced composite part is visualized.

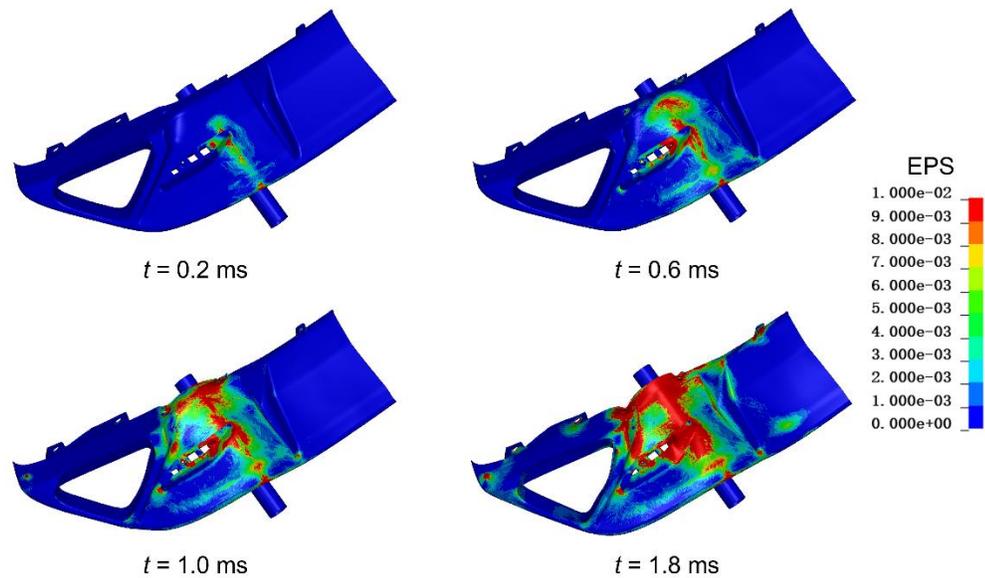

**Fig. 15.** LS-DYNA machine learning-based multiscale simulation results of the homogenized equivalent plastic strain (EPS) distribution on the deformed short-fiber-reinforced composite part during a dynamic impact-contact process.

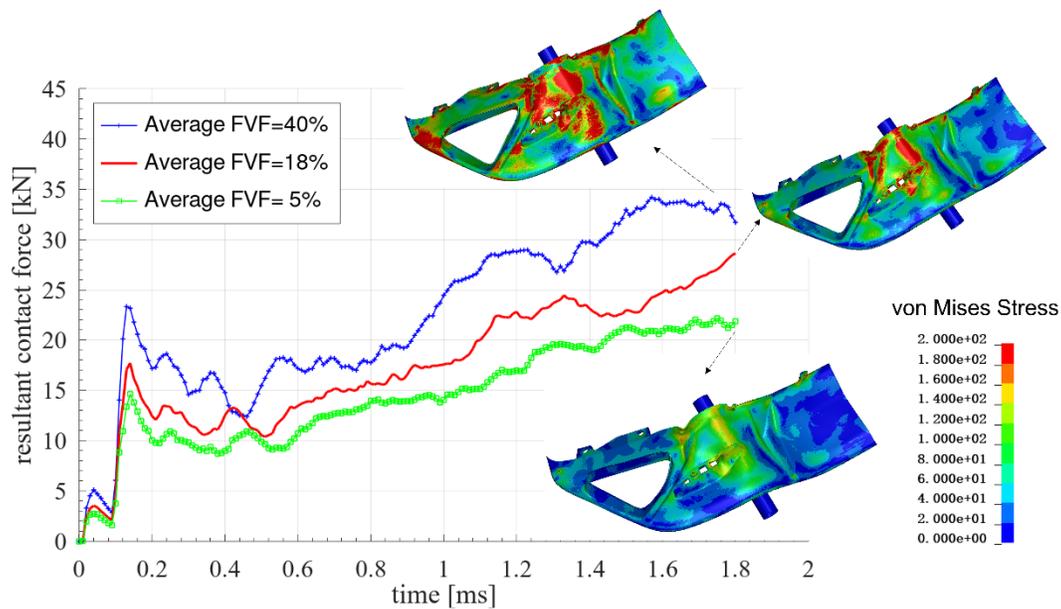

**Fig. 16.** LS-DYNA machine learning-based multiscale simulation results of the resultant contact force and von Mises stress distributions on three short-fiber-reinforced composite parts with 5%, 18%, and 40% average fiber volume fraction (FVF), respectively.

Since the automotive part is produced through injection-molding, the distribution of fiber orientations and volume fractions are influenced by various molding process parameters, such as part shape and thickness, number and position of injection gates, mold temperature, and filling time. In the following, we conducted multiscale structural simulations of the automotive part with different fiber volume fractions to examine the effect of microstructural distribution on the composite mechanical behavior. For illustration purpose, we perform two additional injection molding simulations with a higher fiber weight percentage and a lower fiber weight percentage, respectively. The resulting fiber volume fractions have average values of 5% and 40%, respectively, whereas the predicted fiber orientation distributions are similar to the previous injection molding simulation shown in Fig. 13, as other molding process parameters are kept unchanged. The LS-DYNA multiscale simulation results of these two models are plotted in Fig. 16 together with the result of the previous SFRC model with 18% average fiber volume fraction. As can be seen, SFRC parts with higher fiber volume fractions exhibit stiffer responses undergoing the same dynamic impact process. This shows the effectiveness of the machine learning-based multiscale method for capturing the influence of microstructures on the macroscopic responses, which is crucial for multiscale design and analysis of short-fiber-reinforced composite products. In addition to structural analysis, the present manufacturing process-informed multiscale simulation approach also enables engineers to optimize the molding process design based on the feedback from the mechanical simulation results.

## Conclusions

Injection molding-induced material microstructures cannot be neglected to achieve reliable prediction of the structural responses. Therefore, an effective numerical approach that can capture the effects of local material microstructures (e.g., fiber orientation, fiber volume fraction) on the global composite structural behaviors is of great importance for design and analysis of SFRC structures. In the present work, we have developed an LS-DYNA machine learning-based multiscale method, which is promising for nonlinear modeling of injection-molded SFRC at the industrial scale. A DMN database based on linear elastic data of numerically reconstructed high-fidelity SFRC microstructures is trained in the offline stage. After integrating with finite elements in the engineering simulation software LS-DYNA, new DMN networks corresponding to injection-molding-induced SFRC microstructures are generated online via an efficient transfer learning scheme. The finite element algorithm coupled with DMN is shown to effectively predict the nonlinear material and structural responses of SFRC, and the computation speed is much faster than high-fidelity multiscale finite element models. Since this machine learning model is based on both physics and data, its simulation capability can be continuously enhanced as more high-quality training data are supplied in the future.

To our best knowledge, we have presented the first general purpose finite element analysis package that integrates injection molding-induced microstructures, material homogenization, and mechanistic machine learning for multiscale structural analysis of SFRC. The method has the full potential to be extended for modeling different types of materials, including particle-reinforced composites, continuous fiber-reinforced composites, polycrystalline metals, porous media, etc. We are currently working on these research topics and new results will be reported in future publications.

## Appendix I. Strain Concentration Tensor in DMN

For a two-phase composite, the following equilibrium and kinematic conditions should be satisfied at the material interface:

$$\left(\sigma_{ij}^{p_2} - \sigma_{ij}^{p_1}\right) \cdot n_j = 0 \tag{21}$$

$$u_i^{p_2} - u_i^{p_1} = 0 \tag{22}$$

where the index $i, j = 1, 2, 3$, $\sigma_{ij}^{p_2}$ and $\sigma_{ij}^{p_1}$ denote the component $ij$ of stress tensors for material phases $p_2$ and $p_1$, respectively, $u_i^{p_2}$ and $u_i^{p_1}$ denote the $i^{th}$ displacement component of material phases $p_2$ and $p_1$, respectively, $n_j$ is the component $j$ of a unit normal at the interface. By considering a composite with a two-layer microstructure, where $n_1 = 0$, $n_2 = 0$, $n_3 = 1$, a simple form of the interfacial equilibrium condition can be derived based on Eq. (21):

$$\sigma_{33}^{p_1} = \sigma_{33}^{p_2}, \ \sigma_{23}^{p_1} = \sigma_{23}^{p_2}, \ \sigma_{13}^{p_1} = \sigma_{13}^{p_2} \tag{23}$$

Furthermore, enforcing the kinematic constraint Eq. (22) on the flat interfacial surface (i.e., the *1-2* plane) leads to the following constraints on the strain tensors $\boldsymbol{\varepsilon}^{p_1}$ and $\boldsymbol{\varepsilon}^{p_2}$:

$$\varepsilon_{11}^{p_1} = \varepsilon_{11}^{p_2}, \ \varepsilon_{22}^{p_1} = \varepsilon_{22}^{p_2}, \ \varepsilon_{12}^{p_1} = \varepsilon_{12}^{p_2} \tag{24}$$

Using the Mandel notation, the stress and strain tensors can be converted to the following matrix form:

$$\{\boldsymbol{\sigma}^{p_j}\} = \left\{\sigma_{11}^{p_j} \quad \sigma_{22}^{p_j} \quad \sigma_{33}^{p_j} \quad \sqrt{2}\sigma_{12}^{p_j} \quad \sqrt{2}\sigma_{23}^{p_j} \quad \sqrt{2}\sigma_{31}^{p_j}\right\}^T \tag{25}$$

$$\{\boldsymbol{\varepsilon}^{p_j}\} = \left\{\varepsilon_{11}^{p_j} \quad \varepsilon_{22}^{p_j} \quad \varepsilon_{33}^{p_j} \quad \sqrt{2}\varepsilon_{12}^{p_j} \quad \sqrt{2}\varepsilon_{23}^{p_j} \quad \sqrt{2}\varepsilon_{31}^{p_j}\right\}^T \tag{26}$$

where the superscript $p_j = p_1$ or $p_2$ denotes the material phase. Accordingly, the constitutive model for each material phase can be written as:

$$\{\boldsymbol{\sigma}^{p_j}\} = [\boldsymbol{C}^{p_j}]\{\boldsymbol{\varepsilon}^{p_j}\} \tag{27}$$

Based on the rule of mixture, the averaged strain is defined as

$$\{\bar{\boldsymbol{\varepsilon}}\} = \left(1 - v_f^{p_2}\right)\{\boldsymbol{\varepsilon}^{p_1}\} + v_f^{p_2}\{\boldsymbol{\varepsilon}^{p_2}\} \tag{28}$$

where $v_f^{p_2}$ denotes the volume fraction of material phase $p_2$, which can be calculated from DMN's nodal weights associated with material phases $p_1$ and $p_2$ in the building block. Substituting the kinematic constraint Eq. (24) into Eq. (28), the strain components 11, 22, and 12 of material phase $p_1$ are found to be equal to the averaged strain components:

$$\varepsilon_{11}^{p_1} = \bar{\varepsilon}_{11}, \ \varepsilon_{22}^{p_1} = \bar{\varepsilon}_{22}, \ \varepsilon_{12}^{p_1} = \bar{\varepsilon}_{12} \tag{29}$$

For the remaining strain components, a relationship between the strain of material phase $p_1$ and the averaged strain can be derived by plugging the constitutive Eq. (27) into the interfacial equilibrium Eq. (23) and further considering Eq. (29), which yields the following equation:

$$\begin{Bmatrix} \varepsilon_{33}^{p_1} \\ \sqrt{2}\varepsilon_{23}^{p_1} \\ \sqrt{2}\varepsilon_{31}^{p_1} \end{Bmatrix} = [\hat{\boldsymbol{A}}]\{\bar{\boldsymbol{\varepsilon}}\} \tag{30}$$

in which,

$$[\hat{\boldsymbol{A}}] = \begin{bmatrix} \hat{C}_{33} & \hat{C}_{35} & \hat{C}_{36} \\ \hat{C}_{53} & \hat{C}_{55} & \hat{C}_{56} \\ \hat{C}_{63} & \hat{C}_{65} & \hat{C}_{66} \end{bmatrix}^{-1} [\tilde{\boldsymbol{A}}] \tag{31}$$

$$[\tilde{\boldsymbol{A}}] = \begin{bmatrix} \tilde{C}_{31}^{p_2} & \tilde{C}_{32}^{p_2} & \tilde{C}_{33}^{p_2} & \tilde{C}_{34}^{p_2} & \tilde{C}_{35}^{p_2} & \tilde{C}_{36}^{p_2} \\ \tilde{C}_{51}^{p_2} & \tilde{C}_{52}^{p_2} & \tilde{C}_{53}^{p_2} & \tilde{C}_{54}^{p_2} & \tilde{C}_{55}^{p_2} & \tilde{C}_{56}^{p_2} \\ \tilde{C}_{61}^{p_2} & \tilde{C}_{62}^{p_2} & \tilde{C}_{63}^{p_2} & \tilde{C}_{64}^{p_2} & \tilde{C}_{65}^{p_2} & \tilde{C}_{66}^{p_2} \end{bmatrix} \tag{32}$$

$$\hat{C}_{ij} = \left(1 - v_f^{p_2}\right)C_{ij}^{p_2} + v_f^{p_2}C_{ij}^{p_1} \tag{33}$$

$$\tilde{C}_{ij}^{p_2} = v_f^{p_2}\left(C_{ij}^{p_2} - C_{ij}^{p_1}\right) \tag{34}$$

Combining Eq.(29) and Eq.(30) yields the following equation that relates the DMN building block's average strain to the strain of material phase $p_1$:

$$\{\boldsymbol{\varepsilon}^{p_1}\} = [\boldsymbol{A}]\{\bar{\boldsymbol{\varepsilon}}\} \tag{35}$$

where $[\boldsymbol{A}]$ is the Mandel matrix form of the strain concentration tensor in DMN, and it can be expressed in the following analytical form:

$$[\boldsymbol{A}] = \begin{bmatrix} 1 & 0 & 0 & 0 & 0 & 0 \\ 0 & 1 & 0 & 0 & 0 & 0 \\ \hat{A}_{11} & \hat{A}_{12} & \hat{A}_{13} & \hat{A}_{14} & \hat{A}_{15} & \hat{A}_{16} \\ 0 & 0 & 0 & 1 & 0 & 0 \\ \hat{A}_{21} & \hat{A}_{22} & \hat{A}_{23} & \hat{A}_{24} & \hat{A}_{25} & \hat{A}_{26} \\ \hat{A}_{31} & \hat{A}_{32} & \hat{A}_{33} & \hat{A}_{34} & \hat{A}_{35} & \hat{A}_{36} \end{bmatrix} \tag{36}$$

in which $\hat{A}_{ij}$ is a function of the deep material network's nodal weights and the stiffness of material phases $p_1$ and $p_2$, as defined in Eq. (31).

## Appendix II. Second-Order Fiber Orientation Tensor

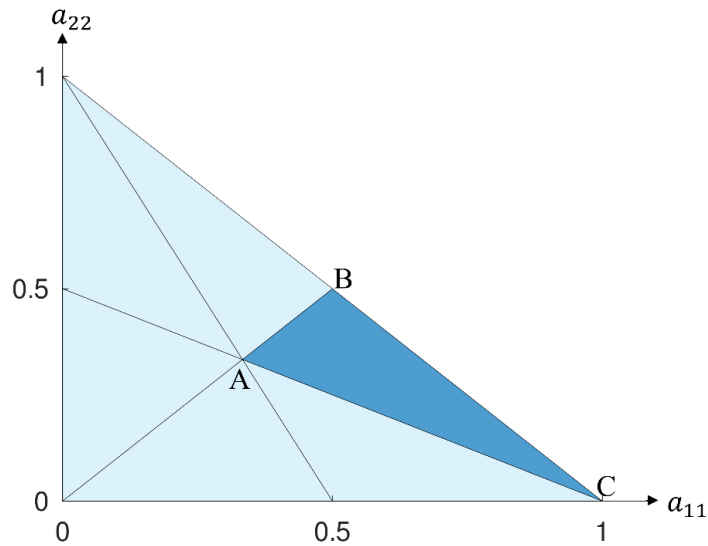

**Fig. 17**. Space of fiber orientation states parametrized by the two largest eigenvalues $a_{11}$ and $a_{22}$ of the second-order fiber orientation tensor $a_{ij}$.

In a short-fiber-reinforced composite part, the fiber orientation state at any spatial point can be represented by the second-order fiber orientation tensor $a_{ij}$ (*Advani and Tucker III, 1987*):

$$a_{ij} = \oint p_i p_j \psi(\boldsymbol{p}) \, d\boldsymbol{p} \tag{37}$$

where $\boldsymbol{p}$ denotes a unit vector along the axial direction of a single short fiber, $\psi(\boldsymbol{p})$ is a probability distribution function, defined so that the probability of finding a fiber oriented within an angular range $d\boldsymbol{p}$ of the direction $\boldsymbol{p}$ is $\psi(\boldsymbol{p})d\boldsymbol{p}$. Since the set of all possible directions of $\boldsymbol{p}$ corresponds to a unit sphere, the integral $\oint p_i p_j \psi(\boldsymbol{p}) \, d\boldsymbol{p}$ over the entire range of $\boldsymbol{p}$ is equivalent to integrating the product of $p_i p_j$ and $\psi(\boldsymbol{p})$ over the surface of a unit sphere. The expression in Eq.(37) clearly shows that the tensor $a_{ij}$ is symmetric. In addition, the normalization condition of the probability distribution function implies that the trace of $a_{ij}$ is always equal to unity. As a result, there are only 5 independent components in $a_{ij}$. If the fiber orientation tensor is rotated into the principal axis system defined by its three orthogonal eigenvectors, it can be expressed in a diagonal matrix form:

$$\boldsymbol{a} = \begin{bmatrix} a_{11} & 0 & 0 \\ 0 & a_{22} & 0 \\ 0 & 0 & a_{33} \end{bmatrix} \tag{38}$$

in which the diagonal components are the three non-negative eigenvalues of $a_{ij}$, and since they satisfy $a_{11} + a_{22} + a_{33} = 1$, there are only 2 independent components left.

If we impose the constraint $a_{11} \geq a_{22} \geq a_{33}$ to the eigenvalues, then all possible fiber orientation states fall into the highlighted triangle A-B-C shown in Fig. 17. Without this constraint, fiber orientation states may exist within other triangular regions in Fig. 17 as well, but all of these orientation states can be mapped onto the highlighted triangular region through rigid body rotations, so we can focus on the smaller region without any loss of generality. As discussed in (*Cintra and Tucker III, 1995*), the vertices of this triangular region have significant physical meanings. The point labeled A corresponds to the random 3D orientation state, where $a_{11} = a_{22} = a_{33} = 1/3$, and all fibers are evenly distributed in all directions. Point B contains the random 2D orientation state, where $a_{11} = a_{22} = 1/2$, $a_{33} = 0$, and all fibers are uniformly distributed in an in-plane direction. Point C corresponds to the unidirectional fiber orientation state, where $a_{11} = 1$, $a_{22} = a_{33} = 0$, and all fibers are parallel to each other. Obviously, a linear combination of fiber orientation tensors from these three vertices can reproduce fiber orientation tensors at any location within the triangle.

## Acknowledgments

The authors would like to acknowledge Zeliang Liu, Tianyu Huang, Dandan Lyu, Yong Guo, Kai Wang, and Philip Ho for their great help with the multiscale method development, and we would also like to thank Madhu Keshavamurthy for continuously supporting this research work. The first author is thankful to Xiaolong He for in-depth discussions on machine learning methods. In addition, we sincerely thank the anonymous reviewers for their constructive comments.

## References

1. Advani, S. G., and C. L. T. I. I. I. 1987. "The Use of Tensors to Describe and Predict Fiber Orientation in Short Fiber Composites." *Journal of Rheology*, 31 (8): 751–784. https://doi.org/10.1122/1.549945.

2. Baek, J., J. S. Chen, and K. Susuki. 2022. "A neural network-enhanced reproducing kernel particle method for modeling strain localization." *International Journal for Numerical Methods in Engineering.* https://doi.org/10.1002/nme.7040.

3. Belytschko, T., W. K. Liu, B. Moran, and K. Elkhodary. 2014. *Nonlinear finite elements for continua and structures.* John Wiley & Sons.

4. Bishara, D., Y. Xie, W. K. Liu, and S. Li. 2022. "A state-of-the-art review on machine learning-based multiscale modeling, simulation, homogenization and design of materials." *Archives of Computational Methods in Engineering*, 1-32. https://doi.org/10.1007/s11831-022-09795-8.

5. Bonatti, C., and D. Mohr. 2022. "On the importance of self-consistency in recurrent neural network models representing elasto-plastic solids." *Journal of the Mechanics and Physics of Solids*, 158, 104697. https://doi.org/10.1016/j.jmps.2021.104697.

6. Bostanabad, R., B. Liang, J. Gao, W. K. Liu, J. Cao, D. Zeng, X. Su, H. Xu, Y. Li, and W. Chen. 2018. "Uncertainty quantification in multiscale simulation of woven fiber composites." *Computer Methods in Applied Mechanics and Engineering*, 338: 506–532. https://doi.org/10.1016/j.cma.2018.04.024.

7. Chen, Z., T. Huang, Y. Shao, Y. Li, H. Xu, K. Avery, D. Zeng, W. Chen, and X. Su. 2018. "Multiscale finite element modeling of sheet molding compound (SMC) composite structure based on stochastic mesostructure reconstruction." *Composite Structures*, 188: 25–38. https://doi.org/10.1016/j.compstruct.2017.12.039.

8. Cintra Jr, J. S., and C. L. Tucker III, 1995. "Orthotropic closure approximations for flow-induced fiber orientation." *Journal of Rheology*, 39(6), 1095-1122. https://doi.org/10.1122/1.550630.

9. Eggersmann, R., T. Kirchdoerfer, S. Reese, L. Stainier, and M. Ortiz. 2019. "Model-free data-driven inelasticity." *Computer Methods in Applied Mechanics and Engineering*, 350: 81–99. https://doi.org/10.1016/j.cma.2019.02.016.

10. Eshelby, J. D., and R. E. Peierls. 1957. "The determination of the elastic field of an ellipsoidal inclusion, and related problems." *Proceedings of the Royal Society of London. Series A. Mathematical and Physical Sciences*, 241 (1226): 376–396. Royal Society. https://doi.org/10.1098/rspa.1957.0133.

11. Feyel, F. 2003. "A multilevel finite element method (FE$^2$) to describe the response of highly non-linear structures using generalized continua." *Computer Methods in Applied Mechanics and Engineering*, 192 (28): 3233–3244. https://doi.org/10.1016/S0045-7825(03)00348-7.

12. Feyel, F., and J.-L. Chaboche. 2000. "FE$^2$ multiscale approach for modelling the

elastoviscoplastic behaviour of long fibre SiC/Ti composite materials." *Computer Methods in Applied Mechanics and Engineering*, 183 (3): 309–330. https://doi.org/10.1016/S0045-7825(99)00224-8.

13. Fish, J., G. J. Wagner, and S. Keten. 2021. "Mesoscopic and multiscale modelling in materials." *Nature Materials*, 20 (6): 774–786. https://doi.org/10.1038/s41563-020-00913-0.

14. Frankel, A. L., R. E. Jones, C. Alleman, and J. A. Templeton. 2019. "Predicting the mechanical response of oligocrystals with deep learning." *Computational Materials Science*, 169: 109099. https://doi.org/10.1016/j.commatsci.2019.109099.

15. Fritzen, F., M. Fernández, and F. Larsson. 2019. "On-the-Fly Adaptivity for Nonlinear Two-scale Simulations Using Artificial Neural Networks and Reduced Order Modeling." *Frontiers in Materials*, 6 (75). https://doi.org/10.3389/fmats.2019.00075.

16. Fritzen, F., and O. Kunc. 2018. "Two-stage data-driven homogenization for nonlinear solids using a reduced order model." *European Journal of Mechanics - A/Solids*, 69: 201–220. https://doi.org/10.1016/j.euromechsol.2017.11.007.

17. Gajek, S., M. Schneider, and T. Böhlke. 2020. "On the micromechanics of deep material networks." *Journal of the Mechanics and Physics of Solids*, 142: 103984. https://doi.org/10.1016/j.jmps.2020.103984.

18. Gajek, S., M. Schneider, and T. Böhlke. 2021. "An FE–DMN method for the multiscale analysis of short fiber reinforced plastic components." *Computer Methods in Applied Mechanics and Engineering*, 384: 113952. https://doi.org/10.1016/j.cma.2021.113952.

19. Gajek, S., M. Schneider, and T. Böhlke. 2022. "An FE-DMN method for the multiscale analysis of thermomechanical composites." *Computational Mechanics*, 69 (5): 1087–1113. https://doi.org/10.1007/s00466-021-02131-0.

20. Gao, J., M. Shakoor, G. Domel, M. Merzkirch, G. Zhou, D. Zeng, X. Su, and W. K. Liu. 2020. "Predictive multiscale modeling for Unidirectional Carbon Fiber Reinforced Polymers." *Composites Science and Technology*, 186: 107922. https://doi.org/10.1016/j.compscitech.2019.107922.

21. Ghaboussi, J., J. H. Garrett, and X. Wu. 1991. "Knowledge-Based Modeling of Material Behavior with Neural Networks." *Journal of Engineering Mechanics*, 117 (1): 132–153. American Society of Civil Engineers. https://doi.org/10.1061/(ASCE)0733-9399(1991)117:1(132).

22. Ghavamian, F., and A. Simone. 2019. "Accelerating multiscale finite element simulations of history-dependent materials using a recurrent neural network." *Computer Methods in Applied Mechanics and Engineering*, 357: 112594. https://doi.org/10.1016/j.cma.2019.112594.

23. Goodfellow, I., Y. Bengio, and A. Courville. 2016. Deep learning. Cambridge, MA: MIT Press.

24. Goury, O., D. Amsallem, S. P. A. Bordas, W. K. Liu, and P. Kerfriden. 2016. "Automatised

selection of load paths to construct reduced-order models in computational damage micromechanics: from dissipation-driven random selection to Bayesian optimization." *Computational Mechanics*, 58 (2): 213–234. https://doi.org/10.1007/s00466-016-1290-2.

25. He, Q., and J. S. Chen. 2020. "A physics-constrained data-driven approach based on locally convex reconstruction for noisy database." *Computer Methods in Applied Mechanics and Engineering*, 363: 112791. https://doi.org/10.1016/j.cma.2019.112791.

26. He, Q., D. W. Laurence, C. H. Lee, and J. S. Chen. 2021a. "Manifold learning based data-driven modeling for soft biological tissues." *Journal of Biomechanics*, 117, 110124. https://doi.org/10.1016/j.jbiomech.2020.110124.

27. He, X., Q. He, and J. S. Chen. 2021b. "Deep autoencoders for physics-constrained data-driven nonlinear materials modeling." *Computer Methods in Applied Mechanics and Engineering*, 385, 114034. https://doi.org/10.1016/j.cma.2021.114034.

28. He, X., Q. He, J. S. Chen, U. Sinha, and S. Sinha. 2020. "Physics-constrained local convexity data-driven modeling of anisotropic nonlinear elastic solids." *Data-Centric Engineering*, 1. E19. https://doi.org/10.1017/dce.2020.20.

29. He, X., and J. S. Chen. 2022. "Thermodynamically Consistent Machine-Learned Internal State Variable Approach for Data-Driven Modeling of Path-Dependent Materials." *Computer Methods in Applied Mechanics and Engineering*, 115348. https://doi.org/10.1016/j.cma.2022.115348.

30. Hessman, P. A., T. Riedel, F. Welschinger, K. Hornberger, and T. Böhlke. 2019. "Microstructural analysis of short glass fiber reinforced thermoplastics based on x-ray micro-computed tomography." *Composites Science and Technology*, 183: 107752. https://doi.org/10.1016/j.compscitech.2019.107752.

31. Huang, T. H., H. Wei, J. S. Chen, and M. C., Hillman. 2020. "RKPM2D: an open-source implementation of nodally integrated reproducing kernel particle method for solving partial differential equations." *Computational Particle Mechanics*, 7(2), 393-433. https://doi.org/10.1007/s40571-019-00272-x.

32. Huang, T., Z. Liu, C. T. Wu, and W. Chen. 2022. "Microstructure-guided deep material network for rapid nonlinear material modeling and uncertainty quantification." *Computer Methods in Applied Mechanics and Engineering*, 398: 115197. https://doi.org/10.1016/j.cma.2022.115197.

33. Huang, W., R. Xu, J. Yang, Q. Huang, and H. Hu. 2021. "Data-driven multiscale simulation of FRP based on material twins." *Composite Structures*, 256: 113013. https://doi.org/10.1016/j.compstruct.2020.113013.

34. Huang, Z. M. 2021. "Constitutive relation, deformation, failure and strength of composites reinforced with continuous/short fibers or particles." *Composite Structures*, 262, 113279. https://doi.org/10.1016/j.compstruct.2020.113279.

35. Ibanez, R., E. Abisset-Chavanne, J. V. Aguado, D. Gonzalez, E. Cueto, and F. Chinesta. 2018. "A manifold learning approach to data-driven computational elasticity and inelasticity." *Archives of Computational Methods in Engineering*, 25(1), 47-57. https://doi.org/10.1007/s11831-016-9197-9.

36. Kaneko, S., H. Wei, Q. He, J. S. Chen, and S. Yoshimura. 2021. "A hyper-reduction computational method for accelerated modeling of thermal cycling-induced plastic deformations." *Journal of the Mechanics and Physics of Solids*, 151, 104385. https://doi.org/10.1016/j.jmps.2021.104385.

37. Karapiperis, K., L. Stainier, M. Ortiz, and J. E. Andrade. 2020. "Data-Driven Multiscale Modeling in Mechanics." *Journal of the Mechanics and Physics of Solids*, 104239. https://doi.org/10.1016/j.jmps.2020.104239.

38. Kirchdoerfer, T., and M. Ortiz. 2016. "Data-driven computational mechanics." *Computer Methods in Applied Mechanics and Engineering*, 304: 81–101. https://doi.org/10.1016/j.cma.2016.02.001.

39. Kochmann, J., S. Wulfinghoff, L. Ehle, J. Mayer, B. Svendsen, and S. Reese. 2018. "Efficient and accurate two-scale FE-FFT-based prediction of the effective material behavior of elasto-viscoplastic polycrystals." *Computational Mechanics*, 61 (6): 751–764. https://doi.org/10.1007/s00466-017-1476-2.

40. Kouznetsova, V. G., M. G. D. Geers, and W. A. M. Brekelmans. 2004. "Multi-scale second-order computational homogenization of multi-phase materials: a nested finite element solution strategy." *Computer Methods in Applied Mechanics and Engineering*, 193 (48): 5525–5550. https://doi.org/10.1016/j.cma.2003.12.073.

41. Le, B. A., J. Yvonnet, and Q.-C. He. 2015. "Computational homogenization of nonlinear elastic materials using neural networks." *International Journal for Numerical Methods in Engineering*, 104 (12): 1061–1084. https://doi.org/10.1002/nme.4953.

42. LeCun, Y., Y. Bengio, and G. Hinton. 2015. "Deep learning." *Nature*, 521 (7553): 436–444. https://doi.org/10.1038/nature14539.

43. Li, S., and G. Wang. 2008. *Introduction to micromechanics and nanomechanics*. World Scientific Publishing Company.

44. Ling, J., R. Jones, and J. Templeton. 2016. "Machine learning strategies for systems with invariance properties." *Journal of Computational Physics*, 318, 22-35. https://doi.org/10.1016/j.jcp.2016.05.003.

45. Liu, X., S. Tian, F. Tao, and W. Yu. 2021. "A review of artificial neural networks in the constitutive modeling of composite materials." *Composites Part B: Engineering*, 224, 109152. https://doi.org/10.1016/j.compositesb.2021.109152.

46. Liu, Z. 2020. "Deep material network with cohesive layers: Multi-stage training and interfacial failure analysis." *Computer Methods in Applied Mechanics and Engineering*, 363: 112913. https://doi.org/10.1016/j.cma.2020.112913.

47. Liu, Z. 2021. "Cell division in deep material networks applied to multiscale strain localization modeling." *Computer Methods in Applied Mechanics and Engineering*, 384: 113914. https://doi.org/10.1016/j.cma.2021.113914.

48. Liu, Z., M. A. Bessa, and W. K. Liu. 2016. "Self-consistent clustering analysis: An efficient multi-scale scheme for inelastic heterogeneous materials." *Computer Methods in Applied Mechanics and Engineering*, 306: 319–341. https://doi.org/10.1016/j.cma.2016.04.004.

49. Liu, Z., M. Fleming, and W. K. Liu. 2018. "Microstructural material database for self-consistent clustering analysis of elastoplastic strain softening materials." *Computer Methods in Applied Mechanics and Engineering*, 330: 547–577. https://doi.org/10.1016/j.cma.2017.11.005.

50. Liu, Z., H. Wei, T. Huang, and C. T. Wu. 2020. "Intelligent multiscale simulation based on process-guided composite database." *16th International LS-DYNA Users Conference,* https://doi.org/10.48550/arXiv.2003.09491.

51. Liu, Z., and C. T. Wu. 2019. "Exploring the 3D architectures of deep material network in data-driven multiscale mechanics." *Journal of the Mechanics and Physics of Solids*, 127: 20–46. https://doi.org/10.1016/j.jmps.2019.03.004.

52. Liu, Z., C. T. Wu, and M. Koishi. 2019a. "A deep material network for multiscale topology learning and accelerated nonlinear modeling of heterogeneous materials." *Computer Methods in Applied Mechanics and Engineering*, 345: 1138–1168. https://doi.org/10.1016/j.cma.2018.09.020.

53. Liu, Z., C. T. Wu, and M. Koishi. 2019b. "Transfer learning of deep material network for seamless structure–property predictions." *Computational Mechanics*, 64 (2): 451–465. https://doi.org/10.1007/s00466-019-01704-4.

54. Lu, X., D. G. Giovanis, J. Yvonnet, V. Papadopoulos, F. Detrez, and J. Bai. 2019. "A data-driven computational homogenization method based on neural networks for the nonlinear anisotropic electrical response of graphene/polymer nanocomposites." *Computational Mechanics*, 64 (2): 307–321. https://doi.org/10.1007/s00466-018-1643-0.

55. Masi, F., I. Stefanou, P. Vannucci, and V. Maffi-Berthier. 2021. "Thermodynamics-based Artificial Neural Networks for constitutive modeling." *Journal of the Mechanics and Physics of Solids*, 147, 104277. https://doi.org/10.1016/j.jmps.2020.104277.

56. Mora-Macías, J., J. Ayensa-Jiménez, E. Reina-Romo, M. H. Doweidar, J. Domínguez, M. Doblaré, and J. A. Sanz-Herrera. 2020. "A multiscale data-driven approach for bone tissue biomechanics." *Computer Methods in Applied Mechanics and Engineering*, 368: 113136. https://doi.org/10.1016/j.cma.2020.113136.

57. Mori, T., and K. Tanaka. 1973. "Average stress in matrix and average elastic energy of materials with misfitting inclusions." *Acta Metallurgica*, 21 (5): 571–574. https://doi.org/10.1016/0001-6160(73)90064-3.

58. Mortazavian, S., and A. Fatemi. 2015. "Effects of fiber orientation and anisotropy on tensile strength and elastic modulus of short fiber reinforced polymer composites." *Composites Part B: Engineering*, 72: 116–129. https://doi.org/10.1016/j.compositesb.2014.11.041.

59. Müller, V., M. Kabel, H. Andrä, and T. Böhlke. 2015. "Homogenization of linear elastic properties of short-fiber reinforced composites–A comparison of mean field and voxel-based methods." *International Journal of Solids and Structures*, 67, 56-70. https://doi.org/10.1016/j.ijsolstr.2015.02.030.

60. Müller, V., and T. Böhlke. 2016. "Prediction of effective elastic properties of fiber reinforced composites using fiber orientation tensors." *Composites Science and Technology*, 130: 36–45. https://doi.org/10.1016/j.compscitech.2016.04.009.

61. Naili, C., I. Doghri, T. Kanit, M. S. Sukiman, A. Aissa-Berraies, and A. Imad. 2020. "Short fiber reinforced composites: Unbiased full-field evaluation of various homogenization methods in elasticity". *Composites Science and Technology*, 187, 107942. https://doi.org/10.1016/j.compscitech.2019.107942.

62. Nemat-Nasser S., and M. Hori. 2013. *Micromechanics: overall properties of heterogeneous materials*. Elsevier.

63. Nguyen, V. D., and L. Noels. 2022a. "Micromechanics-based material networks revisited from the interaction viewpoint; robust and efficient implementation for multi-phase composites." *European Journal of Mechanics - A/Solids*, 91: 104384. https://doi.org/10.1016/j.euromechsol.2021.104384.

64. Nguyen, V. D., and L. Noels. 2022b. "Interaction-based material network: A general framework for (porous) microstructured materials." *Computer Methods in Applied Mechanics and Engineering*, 389: 114300. https://doi.org/10.1016/j.cma.2021.114300.

65. Nordmann, J., M. Aßmus, and H. Altenbach. 2018. "Visualising elastic anisotropy: theoretical background and computational implementation." *Continuum Mechanics and Thermodynamics*, 30(4), 689-708. https://doi.org/10.1007/s00161-018-0635-9.

66. Pasetto, M., J. Baek, J. S. Chen, H. Wei, J. A. Sherburn, M. J. Roth. 2021. "A Lagrangian/semi-Lagrangian coupling approach for accelerated meshfree modelling of extreme deformation problems." *Computer Methods in Applied Mechanics and Engineering*, 381, 113827. https://doi.org/10.1016/j.cma.2021.113827.

67. Rao, C., and Y. Liu. 2020. "Three-dimensional convolutional neural network (3D-CNN) for heterogeneous material homogenization." *Computational Materials Science*, 184, 109850. https://doi.org/10.1016/j.commatsci.2020.109850.

68. Rocha, I. B. C. M., P. Kerfriden, and F. P. van der Meer. 2020. "Micromechanics-based surrogate models for the response of composites: A critical comparison between a classical mesoscale constitutive model, hyper-reduction and neural networks." *European Journal of Mechanics-A/Solids*, 82, 103995. https://doi.org/10.1016/j.euromechsol.2020.103995.

69. Sanz-Herrera, J. A., J. Mora-Macías, J. Ayensa-Jiménez, E. Reina-Romo, M. H. Doweidar, J. Domínguez, and M. Doblaré. 2021. "Data-driven computational simulation in bone mechanics." *Annals of Biomedical Engineering*, 49(1), 407-419. https://doi.org/10.1007/s10439-020-02550-9.

70. Tan, V. B. C., K. Raju, and H. P. Lee. 2020. "Direct FE$^2$ for concurrent multilevel

modelling of heterogeneous structures." *Computer Methods in Applied Mechanics and Engineering*, 360, 112694. https://doi.org/10.1016/j.cma.2019.112694.

71. Tao, F., X. Liu, H. Du, and W. Yu. 2022. "Finite element coupled positive definite deep neural networks mechanics system for constitutive modeling of composites." *Computer Methods in Applied Mechanics and Engineering*, 391, 114548. https://doi.org/10.1016/j.cma.2021.114548.

72. Terada, K., J. Kato, N. Hirayama, T. Inugai, and K. Yamamoto. 2013. "A method of two-scale analysis with micro-macro decoupling scheme: application to hyperelastic composite materials." *Computational Mechanics*, 52(5), 1199-1219. https://doi.org/10.1007/s00466-013-0872-5.

73. Tucker III, C. L., and E. Liang. 1999. "Stiffness predictions for unidirectional short-fiber composites: review and evaluation." *Composites science and technology*, 59(5), 655-671. https://doi.org/10.1016/S0266-3538(98)00120-1.

74. Sanz-Herrera, J. A., J. Mora-Macías, J. Ayensa-Jiménez, E. Reina-Romo, M. H. Doweidar, J. Domínguez, and M. Doblaré. 2021. "Data-Driven Computational Simulation in Bone Mechanics." *Annals of Biomedical Engineering*, 49 (1): 407–419. https://doi.org/10.1007/s10439-020-02550-9.

75. Spahn, J., H. Andrä, M. Kabel, and R. Müller. 2014. "A multiscale approach for modeling progressive damage of composite materials using fast Fourier transforms." *Computer Methods in Applied Mechanics and Engineering*, 268: 871–883. https://doi.org/10.1016/j.cma.2013.10.017.

76. Vlassis, N. N., R. Ma, and W. Sun. 2020. "Geometric deep learning for computational mechanics Part I: anisotropic hyperelasticity." *Computer Methods in Applied Mechanics and Engineering*, 371: 113299. https://doi.org/10.1016/j.cma.2020.113299.

77. Vlassis, N. N., and W. Sun. 2021. "Sobolev training of thermodynamic-informed neural networks for interpretable elasto-plasticity models with level set hardening." *Computer Methods in Applied Mechanics and Engineering*, 377, 113695. https://doi.org/10.1016/j.cma.2021.113695.

78. Vu-Quoc, L., and A. Humer. 2022. "Deep learning applied to computational mechanics: A comprehensive review, state of the art, and the classics." *arXiv Preprint*. https://doi.org/10.48550/arXiv.2212.08989.

79. Wang, H. P., C. T. Wu, Y. Guo, and M. E. Botkin. 2009. "A coupled meshfree/finite element method for automotive crashworthiness simulations." *International Journal of Impact Engineering*, 36(10-11), 1210-1222. https://doi.org/10.1016/j.ijimpeng.2009.03.004.

80. Wang, K., and W. Sun. 2018. "A multiscale multi-permeability poroplasticity model linked by recursive homogenizations and deep learning." *Computer Methods in Applied Mechanics and Engineering*, 334: 337–380. https://doi.org/10.1016/j.cma.2018.01.036.

81. Wang, M. L., R. Y. Chang, and C. H. D. Hsu. 2018. *Molding simulation: Theory and*

*practice.* Carl Hanser Verlag GmbH Co KG.

82. Wei, H., C.T. Wu, D. Lyu, W. Hu, F. H. Rouet, K. Zhang, P. Ho, H. Oura, M. Nishi, T. Naito, and L. Shen. 2021. "Multiscale simulation of short-fiber-reinforced composites: from computational homogenization to mechanistic machine learning in LS-DYNA." *13th European LS-DYNA Conference Proceedings. Ulm, Germany.* https://www.dynalook.com/conferences/13th-european-ls-dyna-conference-2021/composites/wei_ansys_lst.pdf

83. Wei, H., D. Lyu, W. Hu, and C. T. Wu. 2022. "RVE analysis in LS-DYNA for high-fidelity multiscale material modeling." *arXiv Preprint*, https://doi.org/10.48550/arXiv.2210.11761.

84. Wu, C. T., Y. Wu, D. Lyu, X. Pan, and W. Hu. 2020. "The momentum-consistent smoothed particle Galerkin (MC-SPG) method for simulating the extreme thread forming in the flow drill screw-driving process." *Computational Particle Mechanics*, 7: 177-191. https://doi.org/10.1007/s40571-019-00235-2.

85. Wu, L., L. Adam, and L. Noels. 2021. "Micro-mechanics and data-driven based reduced order models for multi-scale analyses of woven composites." *Composite Structures*, 270: 114058. https://doi.org/10.1016/j.compstruct.2021.114058.

86. Xu, K., D. Z. Huang, and E. Darve. 2021. "Learning constitutive relations using symmetric positive definite neural networks." *Journal of Computational Physics*, 428, 110072. https://doi.org/10.1016/j.jcp.2020.110072.

87. Xu, R., J. Yang, W. Yan, Q. Huang, G. Giunta, S. Belouettar, H. Zahrouni, T. ben Zineb, and H. Hu. 2020. "Data-driven multiscale finite element method: From concurrence to separation." *Computer Methods in Applied Mechanics and Engineering*, 363: 112893. https://doi.org/10.1016/j.cma.2020.112893.

88. Yu, C., O. L. Kafka, and W. K. Liu. 2019. "Self-consistent clustering analysis for multiscale modeling at finite strains." *Computer Methods in Applied Mechanics and Engineering*, 349: 339–359. https://doi.org/10.1016/j.cma.2019.02.027.

89. Yvonnet, J., and Q. C. He. 2007. "The reduced model multiscale method (R3M) for the non-linear homogenization of hyperelastic media at finite strains." *Journal of Computational Physics*, 223 (1): 341–368. https://doi.org/10.1016/j.jcp.2006.09.019.

## Data Availability Statement

Some or all data, models, or code generated or used during the study are proprietary or confidential in nature and may only be provided with restrictions.